\documentclass[superscriptaddress,nofootinbib,twocolumn,letterpaper,prb]{revtex4-1}

\usepackage{amsmath,amssymb}
\usepackage{graphicx}
\usepackage[usenames,dvipsnames]{color}
\usepackage{hyperref}
\usepackage{url}
\usepackage[utf8]{inputenc}
\usepackage{slashed}
\usepackage{subfigure}
\usepackage{slashed,bbm}
\usepackage{graphics,psfrag,epsfig}
\usepackage{dsfont}
\usepackage{setspace}

\graphicspath{{figures/}}

\allowdisplaybreaks

\begin{document}

\title{Fermion-induced quantum criticality with two length scales in Dirac systems}

\author{Emilio Torres}
\affiliation{Institute for Theoretical Physics, University of Cologne, 50937 Cologne, Germany}

\author{Laura~Classen}
\affiliation{Physics Department, Brookhaven National Laboratory, Bldg. 510A, Upton, NY 11973, USA}

\author{Igor~F.~Herbut}
\affiliation{Department of Physics, Simon Fraser University, Burnaby, British Columbia, Canada V5A 1S6}

\author{Michael~M.~Scherer}
\affiliation{Institute for Theoretical Physics, University of Cologne, 50937 Cologne, Germany}

\begin{abstract}
The quantum phase transition to a $\mathbb{Z}_3$-ordered Kekul\'e valence bond solid in two-dimensional Dirac semimetals is governed by a fermion-induced quantum critical point, which renders the putatively discontinuous transition continuous. We study the resulting universal critical behavior in terms of a functional RG approach, which gives access to the scaling behavior on the symmetry-broken side of the phase transition, for general dimension and number of Dirac fermions. In particular, we investigate
the emergence of the fermion-induced quantum critical point for space-time dimensions $2<d<4$. We determine the integrated RG flow from the Dirac semi-metal to the symmetry-broken regime and analyze the underlying fixed point structure. We show that the fermion-induced criticality leads to a scaling form with two divergent length scales, due to the breaking of the discrete $\mathbb{Z}_3$ symmetry. This provides another source of scaling corrections, besides the one stemming from being in the proximity to the first order transition.
\end{abstract}

\maketitle

\section{Introduction}

Phase transitions play a pivotal role in our quest for an understanding of the different states of matter.
In fact, the description of many continuous and discontinuous phase transitions in correlated many-body systems is based on a continuum field theory formulation for the order parameter which acquires a non-vanishing expectation value across the transition~\cite{Landau:1980mil}.
Together with the renormalization group (RG) approach\cite{Wegner:1972ih,Wilson:1973jj}, this Landau-Ginzburg-Wilson (LGW) picture not only provides a thorough understanding of the emergence of scaling laws near critical points but in many cases also yields quantitative results for the critical exponents.

With such a successful theory as a basis, it is exciting to ask whether there are critical points in complex many-body systems which go beyond the LGW description.
An exotic scenario which has been discussed in this context are deconfined quantum critical points~\cite{PhysRevB.70.144407,Senthil1490} (DQCP) in which two (ordered) phases are separated by a critical point instead of the expected discontinuity.
It is argued that precisely at the critical point an additional global U(1)~symmetry emerges~\cite{doi:10.1143/JPSJS.74S.1}
and fractional spinon excitations are deconfined, which then govern the critical behavior of the transition.
Furthermore, slightly away from the critical point a second large length scale emerges in addition to the correlation length of order parameter fluctuations.
This is related to the presence of a dangerously irrelevant term in the system and confines the fractionalized degrees of freedom away from the critical point~\cite{doi:10.1143/JPSJS.74S.1}.

In this work, we explore a different scenario for quantum criticality -- the fermion-induced quantum critical points (FIQCPs)~\cite{2015arXiv151207908L,PhysRevB.94.205136,PhysRevB.96.195162,PhysRevB.96.155112,PhysRevB.96.115132} --
which share a number of characteristic properties with DQPCs:
(1)~They exhibit a continuous transition where a discontinuous transition is expected from Landau-Ginzburg theory for the order-parameter field(s).
(2)~They show an emergent U(1) symmetry at the quantum critical point.
(3)~They have two length scales due to the presence of a dangerously irrelevant coupling~\cite{PhysRevB.61.3430,PhysRevB.91.174417,PhysRevLett.115.200601,Shao213}.
FIQCPs are, e.g., relevant to the $\mathbb{Z}_3$-symmetry breaking transition in 3+1D topological Weyl semimetals~\cite{PhysRevB.96.155112} and the semimetal-to-Kekul\'e quantum transition~\cite{PhysRevLett.98.186809,PhysRevB.80.205319,PhysRevB.82.035429} of 2+1 dimensional Dirac fermions on the honeycomb lattice.

The reason for the appearance of critical behavior, despite the expectation of a first order transition~\cite{PhysRevB.8.3419}, is the presence of gapless fermion fluctuations, which cannot be integrated out at zero temperature.
Another essential aspect of this scenario is that it is inherently non-perturbative~\cite{PhysRevB.96.115132}:
The putatively discontinuous transition is caused by an additional canonically relevant coupling and is rendered continuous due to strong fermion fluctuations.
This is in contrast to the conventional LGW picture, where usually only marginally relevant couplings are rendered irrelevant and therefore can be assessed within a perturbative RG approach.
We conclude that a reliable study of the FIQCP scenario requires a non-perturbative theoretical framework which allows to extend the LGW approach in a suitable way.
Such a framework is, for example, provided by the functional renormalization group (FRG), which we will employ here~\cite{Wetterich:1992yh,Berges:2000ew}.

Several previous studies~\cite{2015arXiv151207908L,PhysRevB.94.205136,PhysRevB.96.195162,PhysRevB.96.155112,PhysRevB.96.115132} have investigated the appearance of a FIQCP at the transition to a Kekul\'e order in two-dimensional Dirac systems.
They have mainly focused on aspects (1) and (2) of the aforementioned characteristics of FIQCPs, see Ref.~\onlinecite{PhysRevB.96.195162} for remarks on~(3).
It was found that, indeed, the putatively first order quantum phase transition to a Kekul\'e order in a Dirac system can be of second order.
This, however, depends on the number of Dirac fermions.
Furthermore, it has been shown that a U(1) symmetry emerges at the corresponding fixed point of the RG equations and its related critical exponents have been determined~\cite{2015arXiv151207908L,PhysRevB.96.115132}.

In this work, we extend previous studies with respect to two aspects:
We investigate explicitly the emergence of the second length scale and analyze the fixed point structure including $D<2+1$.
For dimensions close to $D=1+1$, the theory with only order parameter fluctuations has a second order transition, which is in the same universality class as the three-state Potts model~\cite{RevModPhys.54.235}.
There is therefore yet another fixed point, besides the FIQCP, that could potentially be stable and determine the system's behavior.
We show, however, that the Potts fixed point becomes unstable as soon as the gapless Dirac modes are included and that, instead, the Dirac fermions always induce a new critical point (the FIQCP) above a certain critical dimension.

With a FIQCP present, the second length scale is expected to appear because of the non-zero dangerously irrelevant coupling~\cite{PhysRevB.13.2222,AMIT1982207} that describes the Kekul\'e order.
This coupling is related to a cubic order-parameter term,
which is why only a discrete, not a continuous symmetry is broken in the ordered state.
Consequently, there are no Goldstone modes; instead not only the longitudinal but also the transversal mode of the order parameter acquires a mass.
The scaling of the longitudinal mass is as usually related to the relevant coupling of the quantum critical point (QCP).
In contrast, the scaling of the transversal mass has to depend on the scaling of the dangerously irrelevant coupling and defines an additional length scale, which also diverges at the critical point~\cite{PhysRevLett.115.200601}.

To study the appearance of the two length scales, we extend our previous non-perturbative functional RG approach to the symmetry-broken regime.
It allows us to give a comprehensive analysis of the fixed-point structure besides the FIQCP. This includes the Dirac semimetal fixed point, which dominates the long-range behavior of the system in the semi-metallic phase and the Nambu-Goldstone fixed point, which dominates the flow of the system on intermediate scales in the ordered phase.
We calculate improved estimates for the correlation length exponent of the second length scale and show that it is almost identical to the first correlation length exponent provided by the FIQCP.
We therefore suggest that it will be extremely challenging to identify or distinguish this behavior in numerical simulations.
Furthermore, we go beyond a pure fixed-point analysis and calculate the actual flow of the Kekul\'e-Dirac system from the Dirac semi-metal to the Kekul\'e ordered phase.
As a result, we provide a unified picture of the system close to the fermion-induced QCP on both sides of the transition.
%

\paragraph*{Outline.}

We introduce the appropriate Gross-Neveu-Yukawa theory to describe the quantum phase transition of gapless Dirac fermions towards a Kekul\'e valence bond solid (VBS) in Sec.~\ref{sec:GNY}.
In this model the Kekul\'e VBS transition is captured by a complex-valued and $\mathbb{Z}_3$-symmetric order parameter field which is coupled to the fermions.
In Sec.~\ref{sec:FIQCP}, we then explain the systematics of fermion-induced QCPs and recall some of the recent findings and evidence which support this scenario.
Sec.~\ref{sec:FRG}, introduces the non-perturbative functional renormalization group (FRG) method as a suitable approach to describe the symmetry-breaking process occurring at the QCP.
Here, we also explain why it allows to describe the system in the symmetry-broken regime and present our functional RG flow equations.
In Sec.~\ref{sec:results}, we discuss our results for the renormalization group flows in the vicinity of the FIQCP.
In particular, we describe the flow of the system from the symmetric into the symmetry-broken regime which is subject to the influence of various fixed-points.
Technical details on the flow equations are given in the appendix.

\section{Setup}

\subsection{Effective model for the Kekul\'e transition}\label{sec:GNY}

\begin{figure}[t!]
\includegraphics[width=0.8\columnwidth]{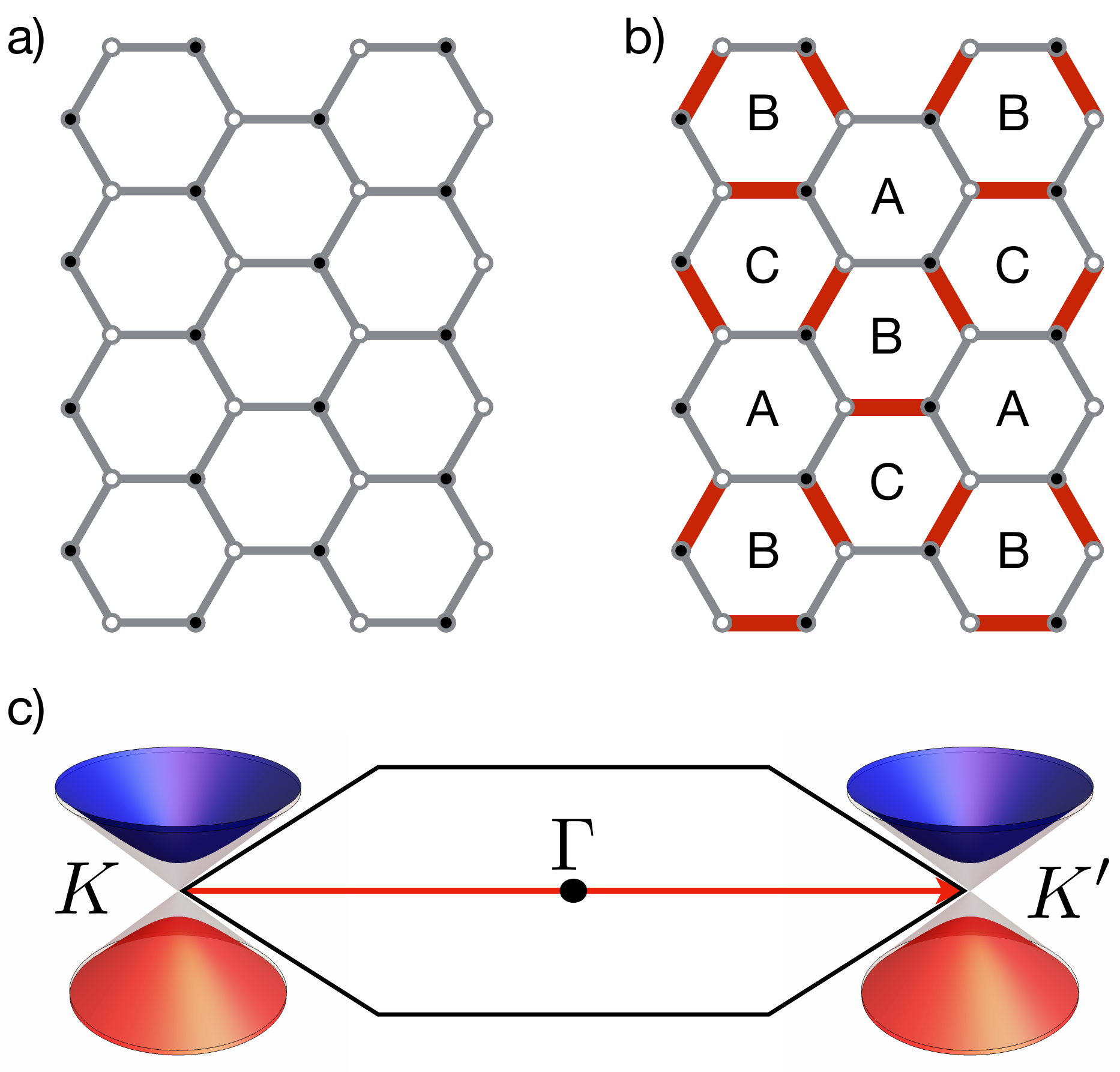}
\caption{{\bf Honeycomb lattice and Kekul\'e dimerization}. (a)~Honeycomb lattice with sublattices $A$ and $B$ represented by the open and filled circles. (b)~Dimerization pattern of the Kekul\'e VBS state, where thick red lines mark stronger bonds and
thinner gray lines mark weaker bonds. (c)~Brillouin zone of the honeycomb
lattice with the Dirac points $K,K^\prime$ where the Kekul\'e pattern opens a mass gap.}
\label{fig:kekule}
\end{figure}

Graphene represents one of the prototypical two-dimensional Dirac semi-metals.
In the simplest model, it~\cite{Novoselov:2005kj} can be described by a tight-binding Hamiltonian of spinless fermions with nearest-neighbor hopping in which the strength of the hopping amplitudes is uniform and real~\cite{RevModPhys.81.109}.
The resulting dispersion has two non-equivalent linear band crossings, the so called Dirac points at $K$ and $K'$.
The low energy physics can be analyzed by restricting the theory to the vicinity of those points.
Different kinds of interactions can, however, open up a gap in the spectrum~\cite{0295-5075-19-8-007,PhysRevLett.97.146401,PhysRevLett.100.146404,PhysRevLett.100.156401,PhysRevB.79.085116,PhysRevB.81.085105,PhysRevB.96.205155}.
One such kind of gap occurs when the nearest-neighbor hopping amplitude develops a modulation with the wave vector connecting the two Dirac points~\cite{PhysRevLett.98.186809,PhysRevB.80.205319,PhysRevB.82.035429,PhysRevB.90.035122,Gomes:2012zza,gutierrez2016imaging}.
In this case a Kekul\'e dimerization pattern occurs, as shown in Fig.~\ref{fig:kekule}.
The angle of the corresponding complex order parameter $\phi=(\phi_1+i\phi_2)/\sqrt{2}$ determines which exact pattern is formed~\cite{PhysRevB.80.205319}.
The transition from the semimetal with its uniform bond strength to the Kekul\'e valence bond solid with modulated bond strength reduces the symmetry of the system:
it no longer possesses the full $C_6$ symmetry of the lattice but only that of its $C_3$ subgroup.
Any of the three patterns of Fig.~\ref{fig:kekule} represents the same transition and the effective theory should be invariant under rotations of $2\pi/3$ of the complex order parameter.

A low-energy effective theory for the system includes the massless four-component fermions.
The four components correspond to the two triangular sublattices, $A$ and $B$, and the two nonequivalent band crossings, $K$ and $K^\prime$.
It also takes into account the dynamics of the complex order parameter $\phi$ and its interaction with the fermions.
The aforementioned terms correspond, respectively (and in that order), to the Lagrangians~\cite{PhysRevLett.53.2449,PhysRevB.82.035429}
\begin{align}
\mathcal{L}_{\psi}&=\sum_{i=1}^{N_f}\overline{\psi}_i\gamma_\mu\partial_\mu\psi_i\;,\label{eq:fermion}\\[5pt]
\mathcal{L}_\phi&=-\lvert\partial\phi\rvert^2+V(\phi,\phi^*)\;,\label{eq:boson}\\[5pt]
\mathcal{L}_{\psi\phi}&=ih\sum_{i=1}^{N_f}\overline{\psi}_i\left(\phi_1\gamma_3+\phi_2\gamma_5\right)\psi_i\;\label{eq:yukawa},
\end{align}
where $N_f$ is the fermion flavor number and $\overline{\psi}:=\psi^\dagger\gamma_0$.
The gamma matrices are given by
\begin{align}
\gamma_0=\mathbbm{1}_2\otimes\sigma_z,&\hspace{0.5cm}\gamma_1=\sigma_z\otimes\sigma_y,\hspace{0.5cm}\gamma_2=\mathbbm{1}_2\otimes\sigma_x\;,\\[5pt]
\gamma_3&=\sigma_x\otimes\sigma_y,\hspace{0.5cm}\gamma_5=\sigma_y\otimes\sigma_y\;, \nonumber
\end{align}
and $V(\phi,\phi^*)$ is a potential for the complex order parameter. The object of our study will thus be the Gross-Neveu-Yukawa theory  with Lagrangian
\begin{align}
\mathcal{L}=\mathcal{L}_{\psi}+\mathcal{L}_{\psi\phi}+\mathcal{L}_{\phi}\;.
\end{align}
We note that $N_f=1$ corresponds to spinless fermions and $N_f=2$ to the case of spin-$1/2$ electrons, as in graphene. Moreover, the Lagrangian for the fermions and the Yukawa interaction alone, cf. Eqs.~\eqref{eq:fermion} and \eqref{eq:yukawa}, is such that the Lagrangian has a $U(1)$ symmetry generated by $\gamma_{35}:=-i\gamma_3\gamma_5$:
\begin{align}
\psi\to e^{i\theta\gamma_{35}/2}\psi\;,\hspace{0.5cm}\phi\to e^{i\theta}\phi\;.
\end{align}
This symmetry is reduced by the potential  $V(\phi,\phi^*)$, which contains all bosonic self-interaction terms allowed by both $U(1)$ and $ \mathbb{Z}_3$ symmetry. Rotationally invariant functions of two variables are functions of $\rho:=\phi^*\phi \propto \phi_1^2+\phi_2^2$,
 while $\mathbb{Z}_3$ invariant functions can in principle be functions of
\begin{align}
\phi^3+\phi^{*3}&=\phi_1^3-3\phi_1\phi_2^2\,,\\
\quad \mathrm{or}\quad i(\phi^3-\phi^{*3})&=\phi^3_2-3\phi_2\phi_1^2\,.
\end{align}
For the specific case of graphene, however, the potential should also be compatible with the discrete symmetries of the system. The Kekul\'e ordering respects both the sublattice symmetry\cite{PhysRevB.80.205319} and the time reversal symmetry. In our setting, their combined action on the order parameter is $\phi_1\to\phi_1$ and $\phi_2\to -\phi_2$.

We therefore conclude that the potential describing the transition is a real function of the two invariants
\begin{align}
\rho&:=\phi^*\phi=\frac{\phi_1^2+\phi_2^2}{2}\;,\label{eq:invariants}\\
\tau^\prime&:=\phi^3+(\phi^*)^3=\frac{\phi_1^3-3\phi_1\phi_2^2}{\sqrt{2}}\;,\nonumber
\end{align}
that is $V(\phi,\phi^*)=U(\rho,\tau')$.
%

\subsection{Fermion-induced quantum critical points}\label{sec:FIQCP}

Since the Kekul\'e VBS is described by a complex-valued order parameter $\phi$ possessing a discrete $\mathbb{Z}_3$ symmetry,
a finite cubic term $\propto \tau^\prime$ is allowed in the free energy $F[\phi]$.
The Landau-Ginzburg mean-field approach then suggests that the minimum of $F[\phi]$ discontinuously jumps from $\phi=0$ to $\phi\neq0$ when the system is tuned through the transition.
This simple assessment, however, underestimates the impact of fluctuations which can become essential for an appropriate description of a system, when the dimensionality is decreased or when additional degrees of freedom are relevant.
In fact, for the Kekul\'e VBS transition in Dirac systems at zero temperature, a description in terms of a $\mathbb{Z}_3$ order parameter field alone appears to be inadequate as the fluctuations of the gapless Dirac fermions can strongly affect the nature of the phase transition.
For instance, it is known that gapless fermion fluctuations near a transition with an O(N)-symmetric order parameter modify the critical behavior and define the chiral universality classes~\cite{Gat:1991bf,Rosenstein:1993zf,PhysRevB.96.165133,Iliesiu:2017nrv,PhysRevD.96.096010}.
Moreover, Refs.~\onlinecite{2015arXiv151207908L,PhysRevB.94.205136,PhysRevB.96.195162,PhysRevB.96.155112,PhysRevB.96.115132} find evidence that the expected discontinuous Kekul\'e transition is rendered continuous in 2+1 dimensions.

From a RG point of view, the change of the phase transition from discontinuous to continuous can be rationalized by considering the scaling dimensions of the cubic order-parameter terms.
At the microscopic level, their presence suggests a second RG relevant direction with sizable power-counting dimension $[g]=3-D/2$ with spacetime dimension $D$.
At a non-trivial fixed point, fluctuation corrections modify the RG scaling of parameters as compared to the scaling suggested by dimensional analysis~\cite{PhysRevLett.28.240,ZinnJustin:1999bf}.
The order of the phase transition then depends on whether the direction corresponding to the cubic coupling remains relevant or becomes irrelevant at the fixed point.
This means when the fluctuations from the gapless Dirac fermions are strong enough to render the canonically relevant direction from the cubic coupling irrelevant, a continuous transition is induced:
Only one relevant direction remains, i.e. a single tuning parameter is sufficient to drive the system to the non-trivial fixed point and universal critical behavior can be observed~\cite{PhysRevB.96.115132}.

Whether or not fluctuations change the transition to the $\mathbb{Z}_3$ ordered state from discontinuous to continuous is evidently a non-perturbative problem, as it systematically requires strong fluctuations.
Therefore, non-perturbative methods are needed to reliably describe this scenario of a FIQCP.
Evidence for the validity of this scenario has been gathered by two complementary non-perturbative approaches, a lattice QMC study~\cite{2015arXiv151207908L} and in terms of the FRG~\cite{PhysRevB.96.115132}.
More specifically, the FRG approach studied a comprehensive regime of the number of Dirac fermions $N_f$ to find a continuous transition in $D=2+1$ dimensions if $N_f>N_{f,c}\approx1.9$.
This was found to be in agreement with the limits provided by QMC~\cite{2015arXiv151207908L}  ($N_{f,c}<2$) and SUSY~\cite{PhysRevB.96.195162,Zerf:2016fti,PhysRevD.96.096010}  ($N_{f,c}>1/2$) calculations and is relevant to the case of spin-1/2 fermions on the honeycomb lattice, $N_f=2$.

\section{FRG method and flow equations}\label{sec:FRG}

\subsection{Method}

The functional renormalization group is a reformulation of Wilson's idea of evaluating the partition function $\mathcal{Z}$ by integrating out fluctuations step by step.
Decoupling fast and slow contributions to the partition function at the scale $k$ is implemented by the insertion of a mass-like regulator function $R_k(q)$  into the bare action $S$ that suppresses fluctuations of field modes with $|q|<k$.

Explicitly, if $\Phi$ contains all the fields of the problem, the partition function at scale $k$ is defined as $\mathcal{Z}_k:=\int D\Phi e^{-S_k[\Phi]}$, with $S_k=S+\frac{1}{2}\int_q\Phi(-q)R_k(q)\Phi(q)$.
The regulator is chosen such that $R_k(q)\to 0$ when $k\to0$, and $R_k(q)\to\infty$ for $k\to\Lambda$, where $\Lambda$ is the inverse lattice spacing, i.e. the highest energy scale of the system.
By the choice of regulator, the (modified) Legendre transform of $\mathcal{Z}_k$, $\Gamma_k$, is such that $\Gamma_{k\to\infty}=S$ and $\Gamma_{k=0}=\Gamma$ is the full quantum effective action.

Setting $t=\ln(k/\Lambda)$, the evolution of $\Gamma_k$ is determined by the Wetterich equation \cite{Wetterich:1992yh,Berges:2000ew}:
\begin{align}
\partial_t\Gamma_k=\frac{1}{2}\mbox{Str}\left[(\Gamma^{(2)}_k+R_k)^{-1}\partial_tR_k\right]\;,
\label{eq:wetteq}
\end{align}
where $\Gamma^{(2)}_k$ is the Hessian of $\Gamma_k$, i.e.,
\begin{align}
\Big(\Gamma_k^{(2)}\Big)(p,q)=\frac{\overrightarrow{\delta}}{\delta\Phi(-p)^T}\Gamma_k\frac{\overleftarrow{\delta}}{\delta\Phi(q)}\,.
\end{align}
The key features of the FRG approach in the present context are the following:
The FRG is well-suited for calculations directly in 2+1 dimensions. It has been tested against other methods which has shown that it provides very good results in the context of Gross-NeveuYukawa models\cite{Rosa:2000ju,Hofling:2002hj,PhysRevD.81.025009,Janssen:2012pq,Mesterhazy:2012ei,Janssen:2014gea,PhysRevB.93.125119,Gehring:2015vja,PhysRevD.93.125021,Knorr:2016sfs,Knorr:2017yze,Yin:2017gkv,Gies:2017tod,PhysRevB.97.041117,Feldmann:2017ooy}.
Importantly, the FRG approach already includes resummation effects through the non-perturbative threshold functions\cite{delamotte2005can,PhysRevB.82.104432}, which is in contrast to perturbative approaches.
Furthermore, it allows a continuation of  renormalization group flows into the symmetry-broken regime and an extraction of  physical information about continuous as well as discontinuous transitions\cite{PhysRevLett.84.5208,Schaefer:2004en,Braun:2007bx,PhysRevLett.103.220602,Pawlowski:2014zaa}.

\subsection{Truncations}

Faced with the impossibility of solving equation \eqref{eq:wetteq} exactly, we employ an ansatz for $\Gamma_k$, inspired by the original form of $S$, that is
\begin{align}
	\Gamma_k&=\int d^Dx\Big[
	Z_{\psi,k}\overline{\psi}\gamma_\mu \partial_\mu\psi+i h_k\overline{\psi}\left(\phi_1\gamma_3+\phi_2\gamma_5\right)\psi\nonumber\\[5pt]
	&\hspace{0.9cm}-\frac{1}{2}Z_{\phi,k}(\phi_1\partial_\mu^2\phi_1+\phi_2\partial_\mu^2{\phi}_2)+U_k(\rho,\tau^\prime)\Big]\label{eq:effaction}\,,
\end{align}
where $Z_{\psi,k}$ and $Z_{\phi,k}$ are the running wavefunction renormalization constants, $\rho,\tau^\prime$ are the bosonic field invariants defined in Eq.~\eqref{eq:invariants} and \eqref{eq:newtau}, and $h_k$ is the running Yukawa coupling. To simplify notation, we defined $\psi:=\oplus_i\psi_i$, so that our gamma matrices are now of size $N_f d_\gamma$, with $d_\gamma$ the dimension of the chosen representation of the Clifford algebra of $\mathbb{R}^{2+1}$. The potential $U_k(\rho,\tau^\prime)$ is defined as $\Gamma_k[\phi,\phi^*]=\Omega\, U_k(\rho,\tau^\prime)$, where the fields $\phi,\phi^*,\psi$ are held constant and where $\Omega$ is the volume of the system.

In the symmetry-broken regime, the potential $V(\phi,\phi^\ast)$ exhibits three equivalent minima.
The physics is generally determined by the global minimum of the potential and its surrounding. Consequently, we use an expansion around one of these three minima in the following. To facilitate the resulting expressions, we introduce another $\mathbb{Z}_3$-invariant quantity for this, which is a combination of $\rho$ and $\tau^\prime$:
\begin{align}
\tau:=\tau'+2\rho^{3/2}\;.
\label{eq:newtau}
\end{align}
With this definition, we find that $\tau=0$ when evaluated at a minimum of $V(\phi,\phi^\ast)$. We note that the whole procedure could also be carried out in terms of monomials of the original fields.

To study the FIQCP from both sides of the transition we now expand the potential $U_k$ around $\rho=\tau=0$ for the symmetric phase and around one of its running minima, $\kappa_{\rho,k}:=\rho^{\min}_{j,k}\neq 0$ for the symmetry broken regime. Here, $j\in \{1,2,3\}$ enumerates the three equivalent minima of the potential.
The two expansions of the potential, then explicitly read
\begin{align}
U_k(\rho,\tau)&=\sum_{m+n=1}^{2m+3n=N}\frac{\lambda_{m,n;k}}{m!n!}\rho^m\tau^n\;,\\[5pt]
U_k(\rho,\tau)&=\Lambda_{0,1;k}\tau\notag\\
&\quad+\sum_{m+n=2}^{2m+3n=N}\frac{\Lambda_{m,n;k}}{m!n!}(\rho-\kappa_{\rho,k})^m\tau^n\;,
\label{eq:potentialSSB}
\end{align}
where
\begin{align}
\lambda_{m,n;k}&:=\frac{\partial^{m+n}U_k}{\partial\rho^m\partial\tau^n}\Big\lvert_{\rho=\tau=0}\,,\\[5pt]
\Lambda_{m,n;k}&:=\frac{\partial^{m+n}U_k}{\partial\rho^m\partial\tau^n}\Big\lvert_{\rho=\kappa_\rho,\tau=0}\;,\label{eq:boskopp1}\\[5pt]
0&=\frac{\partial U_k}{\partial \rho}\Big\lvert_{\rho=\kappa_\rho,\tau=0}\,,
\label{eq:boskopp2}
\end{align}
and $N$ denotes the order of the truncation.
The truncation employed, here, is referred to as the extended local potential approximation of order $N$, or LPA$N^\prime$, for short. We use the freedom in choosing any of the three minima by choosing the one in which $\tau$ vanishes, which is equivalent to the choice $\phi_2^{\min}=0$.

\subsection{Flow equations}

Flow equations for the bosonic couplings are obtained by acting with $\partial_t$ on both sides of Eqs. \eqref{eq:boskopp1} and \eqref{eq:boskopp2}, while those for the Yukawa coupling and the anomalous dimensions are obtained from the projections
\begin{align}
h_k&=\frac{-i}{N_f d_\gamma}\text{Tr}\left[ \gamma_5 \frac{\delta}{\delta \Delta \phi_2(p')} \frac{\delta}{\delta \bar \Psi(p)}\Gamma_k \frac{\delta}{\delta \Psi(q)}\right]\,,\\
Z_{\phi,k}&=\frac{\partial}{\partial p^2}\int_q \frac{\delta}{\delta\phi_2(-p)} \frac{\delta}{\delta\phi_2(q)} \Gamma_k \\
Z_{\psi,k}&=\frac{-i}{N_fd_\gamma D}\text{Tr}\left[\gamma_\mu\frac{\partial}{\partial p_\mu}\int_q \frac{\delta}{\delta\bar\psi(p)}\Gamma_k \frac{\delta}{\delta\psi(q)}\right]\;.
\label{eq:projections}
\end{align}
All of the above expressions are evaluated at the minimum of the order-parameter potential and zero momenta $p=q=0$, and the definitions
\begin{align}
\eta_\phi=-\partial_t\log Z_{\phi,k}\,,\hspace{0.5cm}\eta_{\psi}=-\partial_t\log Z_{\psi,k}\;.
\label{eq:anomdims}
\end{align}
To discuss fixed point properties, it is necessary to switch to dimensionless variables (denoted by a bar), defined as
\begin{align}
\overline{\rho}=\frac{k^{D-2}}{Z_{\phi,k}}\rho\,,\quad \mathrm{and}\quad \overline{\tau}=\frac{k^{3(D-2)/2}}{Z_{\phi,k}^{3/2}}\tau\,.
\end{align}
The computation of the flow requires inversion of the Hessian, which we do by computing its spectrum, see Eq.~\eqref{eq:wetteq}.
The eigenvalues in the bosonic sector, obtained from the symmetry broken expansion \eqref{eq:potentialSSB}, define the longitudinal and transverse masses $\overline{m}_L$ and $\overline{m}_T$. In terms of the invariants defined in \eqref{eq:invariants} and \eqref{eq:newtau}, the masses take a complicated form, see Eq.~\eqref{eq:massesfull}in App.~\ref{app:flow}.
These expressions simplify when evaluated at the minimum of our choice to
\begin{align}
\overline{m}_{L,\min}^2&=2\overline{\kappa}_\rho\Lambda_{2,0}\;,\label{eq:massL}\\
\overline{m}_{T,\min}^2&=9\sqrt{\overline{\kappa}_\rho}\Lambda_{0,1}\;,
\label{eq:massT}
\end{align}
and, as expected, there are no Goldstone modes in the system for any nontrivial dependence of $U$ on $\tau$ since only a discrete symmetry is broken.
In other terms, the masses $\overline{m}_L,\overline{m}_T$ are always different and nonzero except exactly at the phase transition.
The appearance of a nonzero transversal mass is the reason behind the emergence of a second length scale in the symmetry-broken regime. Note, moreover, that this phenomenon can only be seen when $\kappa_\rho\neq0$ and, in particular, observing its effect depends crucially on the fact that we can follow the evolution of the system for arbitrarily large values of $\kappa_\rho$, which is not a feasible task in a perturbative approach.
In terms of these masses, the flow equation for the dimensionless potential $u=k^{-D}U$ takes the form
\begin{align}
\partial_tu&=-Du+\frac{1}{2}(D-2+\eta_\phi)\left(2\overline{\rho} u^{(1,0)}+3\overline{\tau} u^{(0,1)}\right)\nonumber\\
&+2v_D\left(l_B(\overline{m}_L^2)+l_B(\overline{m}_T^2)\right)-2v_DN_fd_\gamma l_F(\omega_\psi).
\label{eq:flowU}
\end{align}
For the Yukawa coupling $\bar{h}^2=k^{D-4}Z_{\phi,k}^{-1}Z_{\psi,k}^{-2}h^2$ it reads
\begin{align}
\partial_t \bar{h}^2=&(D-4+\eta_\phi+2\eta_\psi)\bar{h}^2-8v_D \bar{h}^4 \left(l_{11}^{FR_2} - l_{11}^{FR_1} \right)\nonumber\\
&-16v_D\sqrt{2\bar\kappa_\rho} \bar{h}^4 u_{122}l_{111}^{FR_1R_2}\;.
\label{eq:yukawaflow}
\end{align}
Additionally, the anomalous dimensions are given by
\begin{align}
\eta_\psi=& \frac{8v_D}{D}\bar{h}^2 \left( m_{(12) R_1}^{FB} + m_{(12) R_2}^{FB} \right)\;,\label{eq:anomdimsflow1}\\
\eta_\phi=&\frac{4v_D}{D}\Big[ m_{4R_2}^B(u_{222},u_{221}) + m_{4R_1}^B(u_{211},u_{221})\nonumber\\
&+ 2m_{(22)R_1R_2}^B(u_{221},u_{211},u_{222})\label{eq:anomdimsflow2}\\
&+ 2N_fd_\gamma \bar{h}^2 \left( m_{4}^F(\omega_\psi) + 2\bar{h}^2\bar\kappa_\rho m_{2}^F(\omega_\psi) \right) \Big]\;,\nonumber
\end{align}
where $v_D=(2^{D+1}\pi^{D/2}\Gamma(D/2))^{-1}$. The different threshold functions $l_{\cdots}^{\cdots},m_{\cdots}^{\cdots}$ and arguments $u_{ijk}$ are listed in App.~\ref{app:flow}.
For a given truncation of the potential including powers of the field up to $N$, Eqs.~\eqref{eq:flowU} to \eqref{eq:anomdimsflow2} form a closed set of coupled differential equations, cf. Ref.~\onlinecite{PhysRevB.96.115132}.

\section{Results}\label{sec:results}

The FRG approach as set up in the previous section allows us to study the fixed-point properties of the system, as well as the RG flows in the symmetric and the symmetry-broken regime.
In this section, we first extend previous investigations of the FIQCP by including spacetime dimensions $1+1\leq D\leq 2+1$ which are accessible with the FRG~\cite{PhysRevLett.75.378,PhysRevB.64.054513,Codello:2012sc,PhysRevLett.110.141601,Janssen:2014gea,Borchardt:2016kco}.
Then, we present how the system flows from the Dirac semimetal regime to the ordered phase and we explain the different regimes of this flow in terms of the characteristic fixed point structure. Finally, we give estimates for the scaling exponent of the second length scale that emerges in the ordered phase.

\subsection{Fermion-induced QCP below $D=2+1$}

\begin{figure}[h!]
\includegraphics[width=0.9\columnwidth]{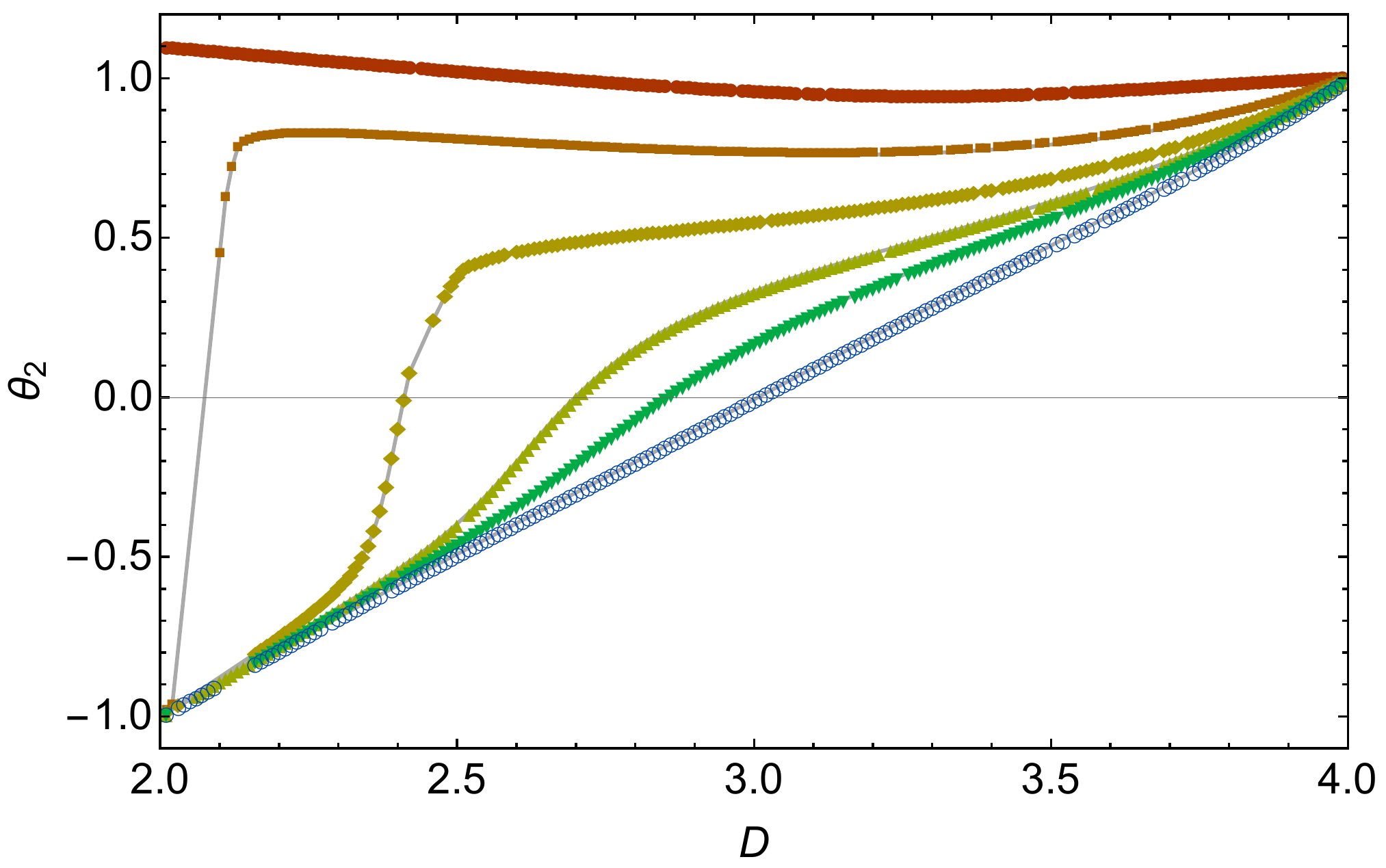}
\caption{{\bf Stability exponent $\theta_2$} for the $U(1)$ symmetric non-Gau\ss ian fixed point of the $\mathbb{Z}_3$ GNY model. In the purely bosonic limit, this corresponds to the O(2)-symmetric Wilson-Fisher FP. We show different small numbers of fermion flavors $N_f$ as function of the dimension. From top to bottom, we show $N_f \in \{0,\frac{1}{4},\frac{1}{2},\frac{3}{4},1,2\}$. As soon as fermions are added, this FP becomes stable below some critical dimension. The results have been calculated in LPA8${}^\prime$.}
\label{fig:WFevol}
\end{figure}

In our previous work~\cite{PhysRevB.96.115132}, we analyzed FRG fixed points in the symmetric regime.
We showed that the FIQCP appears above a critical $N_{f,c}\approx1.9$ in $D=2+1$.
Moreover, the FIQCP is characterized by an emergent U(1) symmetry where all couplings that break the U(1) down to $\mathbb{Z}_3$ vanish at the fixed point. Therefore, the FIQCP coordinates and, more importantly, a subset of the critical exponents coincide with the ones from the chiral XY model which also exhibits a global O(2) $\cong$ U(1).

Here, we show that this is also true for lower dimensions, cf. also Ref.~\onlinecite{PhysRevB.96.195162}.
We find that for any given non-zero number of Dirac fermions $N_f$, there is a critical dimension below which an O(2) symmetric Gross-Neveu-Yukawa fixed point becomes stable so that a second order transition is induced realizing the FIQCP scenario.
This can be seen in Fig.~\ref{fig:WFevol}, where we plot the second largest critical exponent $\theta_2$ of the O(2)-symmetric fixed point for different dimensions and $N_f$.
If $\theta_2<0$, the fixed point is stable and describes a second order phase transition. We see how $\theta_2$ of the O(2)-symmetric FP changes from the case without fermions $N_f=0$ to the one with fermions and that it drops below zero at a critical dimensions $D_c>2$ as soon as $N_f\neq0$. This critical dimension continuously connects to the value that was found before~\cite{PhysRevB.96.115132}.

Interestingly, there is potentially another fixed point that can yield a second order transition for dimensions close to $D=2$: in the system without fermions, this fixed point corresponds to the phase transition of the three-state Potts model~\cite{Baxter:2000ez,RevModPhys.54.235,Zinati:2017hdy}, and disappears above a certain critical dimension in the vicinity of $D=3$, see App.~\ref{app:potts}.
We find, however, that as soon as the fermions are included, this Pott's fixed point always becomes unstable, so that the O(2)-symmetric FIQCP is the only possibility to obtain a second order transition.
The reason for the destabilization of the Potts fixed point upon inclusion of fermions is that, even for small $N_f$, we introduce another RG direction represented by the Yukawa coupling $\overline{h}$. 
At the Potts fixed point, the Yukawa coupling is $\overline{h}^\ast=0$, and below $D=4$, this always introduces a relevant direction to the Potts fixed point making it unstable.
%

\subsection{Flow from Dirac semimetal to Kekul\'e order}

Turning back to the physical case of $D=2+1$, we now study the renormalization group flow of the model, which exhibits a rich structure.
In the phase diagram of the considered Gross-Neveu-Yukawa model, the FIQCP separates the symmetric or Dirac-semimetal (DSM) from the symmetry-broken regime. To see the scaling behavior as induced by the FIQCP, a fine-tuning of the RG-relevant parameter is required. Eventually, in the deep infrared, when almost all momentum-modes have been integrated out, the system ends up either in the DSM phase or in the symmetry-broken phase.
To understand the semimetal-to-Kekul\'e-VBS transition in terms of the renormalization group flow, we have to consider the fixed point structure beyond the FIQCP.
Generally, the renormalization group flow to the symmetry-broken phase in the vicinity of the fermion-induced QCP proceeds as follows~\cite{PhysRevB.96.195162}:

\begin{figure}[b!]
\includegraphics[width=0.8\columnwidth]{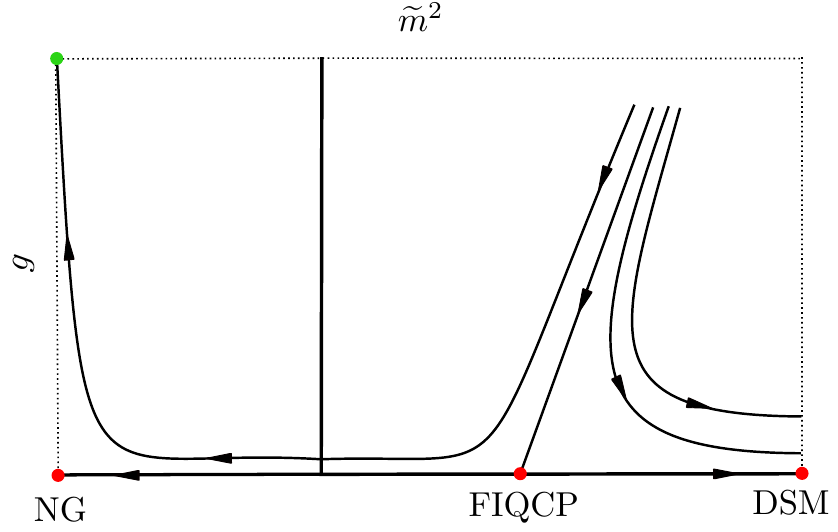}
   \caption{{\bf Schematic flow of the cubic coupling} as a function of a tuning parameter $\widetilde{m}^2$. $\widetilde{m}^2>0$ indicates the system is in the symmetric regime and $\widetilde{m}^2<0$ that it is in the symmetry broken regime. Vertical dotted lines are $\widetilde{m}^2=\pm\infty$. We show four trajectories for identical initial $g$ and different values of~$\widetilde{m}^2$.}
   \label{fig:flowscheme}
\end{figure}

(1)~At microscopic scales, the RG flow of the system is initialized in the symmetric regime and a fine-tuning of the mass parameter $\overline{m}^2$ has to be performed to drive the system close to the FIQCP.

(2)~The system still remains in the symmetric regime on intermediate scales  where it approaches the FIQCP which then dominates the scaling behavior. In this regime the cubic coupling, $\lambda_{0,1}=:g$, is attracted to its fixed-point value $g^\ast=0$ and therefore becomes small.

(3)~After some RG-time close to the FIQCP, the flow  departs from it and goes towards the fixed point that characterizes the symmetry-broken phase of the O(2) model - the Nambu-Goldstone (NG) fixed point. We note that due to the small but finite cubic coupling, the transversal mode already acquires a small mass.

(4) Finally, as the cubic coupling is relevant at the NG fixed point and $g\neq 0$ the flow will be driven away from the NG fixed point and the mass of the pseudo-Goldstone mode becomes more pronounced which is related to the appearance of the second length scale $\xi^\prime$.

\begin{figure}[t!]
\includegraphics[width=\columnwidth]{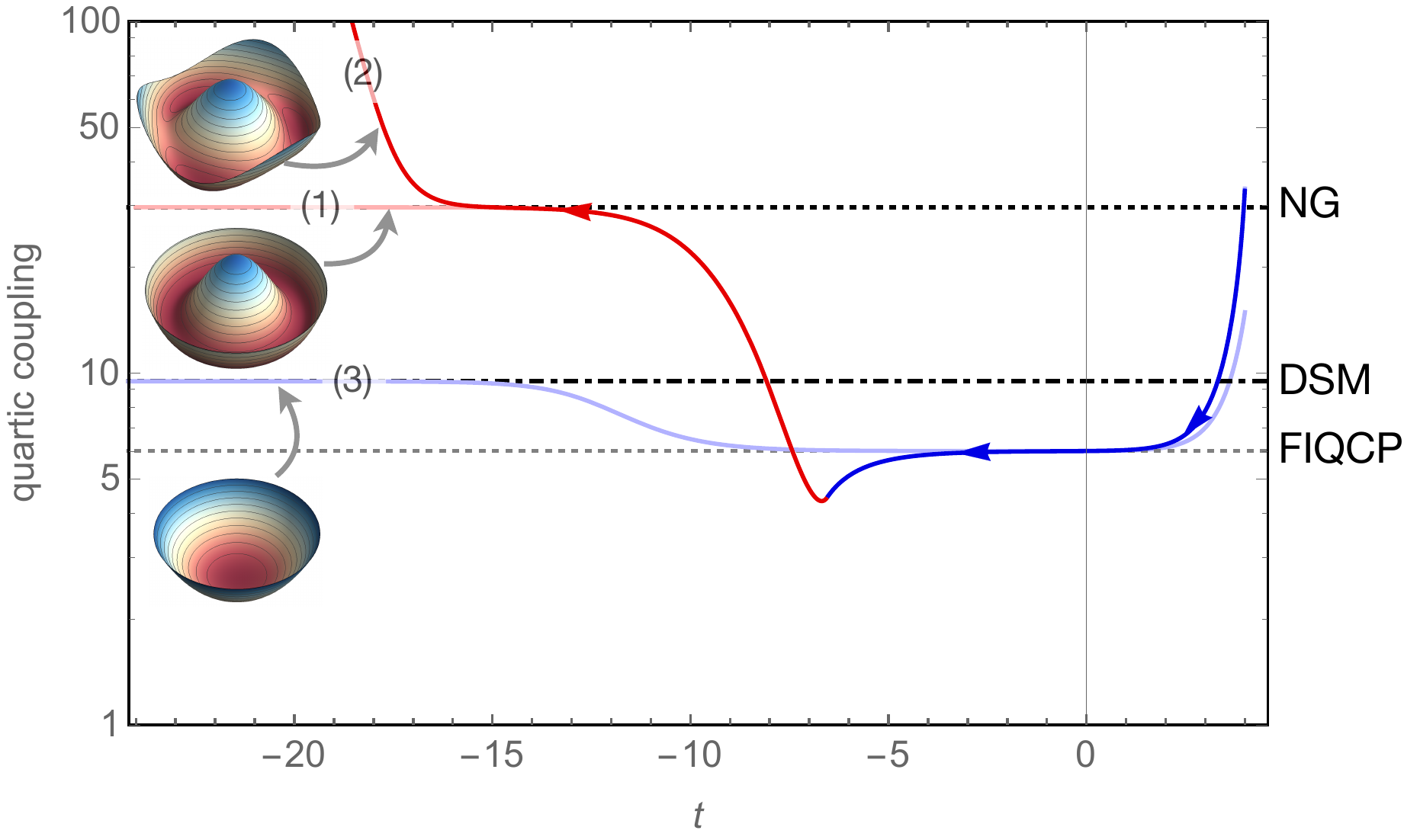}
\caption{{\bf Flow of the quartic coupling} into the different regimes: (1)~for $g\equiv 0$ the flow generically starts in the symmetric regime in the ultraviolet and can be fine-tuned to approach the FIQCP. Later, it may enter the SSB regime (as indicated by the change of color from blue to red) and flows to the NG fixed point, where it remains (light red line). (2)~For small $g\neq 0$ the flow trajectory is almost identical during the entire flow. Only in the deep IR it departs from the NG fixed point due to the dangerously irrelevant direction corresponding to the cubic operator. (3)~There can also be flows which remain completely in the symmetric regime and no symmetry breaking occurs. In this case the flow approaches the DSM fixed point in the deep IR as indicated by the light blue line.
In the insets, we schematically show the shape of the effective potential in the different regimes.}
\label{fig:flowgen}
\end{figure}

When the initial system is not fine-tuned, the flow will stay in the symmetric regime, dominated by a fixed point that we call DSM fixed point in the following.
This behavior is schematically shown in Fig.~\ref{fig:flowscheme}.
In the following, we describe the different fixed point regimes characterizing the symmetric and ordered phase in detail.
Furthermore, we show representative RG flow trajectories of the quartic coupling in Fig.~\ref{fig:flowgen} running through all the described regimes explicitly.
Thereby the different fixed point regimes are revealed as plateaus where the flow of the quartic coupling hardly changes. However, if there is a relevant direction, the flow is eventually driven away from the corresponding fixed point value.

\subsubsection{Dirac semimetal regime.}

When the system is in the DSM phase it is eventually dominated by the symmetric DSM fixed point which can be analytically found in the symmetric FRG flow equations upon taking the limit $\lambda_{1,0}=:\overline{m}^2\to\infty$. In this limit, we obtain the following $\beta$ functions for the Yukawa coupling ($\overline{h}^2$), the cubic ($g$) and the quartic scalar coupling ($\lambda:=\lambda_{2,0}$) within LPA${^\prime}4$
\begin{align}
\beta_{\overline{h}^2}&=\overline{h}^2\left(D-4+2c_DN_f\overline{h}^2 \frac{3D-4}{D(D-2)}\right)\,, \\[5pt]
\beta_{g}&=\frac{g}{2}\left(D-6+6 c_D N_f \overline{h}^2\frac{3D-4}{D(D-2)}\right)\,, \\[5pt]
\beta_{\lambda}&=\lambda(D-4)-\frac{4c_D N_f \overline{h}^2}{D}\left(8\overline{h}^2+\lambda\frac{4-3D}{(D-2)}\right)\,,
\end{align}
where $c_D=d_\gamma v_D$.

We note that these $\beta$ functions are not restricted to a finite expansion in the LPA${^\prime}$, but are valid to any order, i.e. in particular for LPA$n^\prime$ with $n\to \infty$.
They admit a nontrivial fixed-point solution
\begin{align}
\overline{h}^{2,\ast}_{\mathrm{DSM}}&=\frac{D}{2c_DN_f}\frac{(4-D)(D-2)}{(3D-4)}\,,\label{eq:h2DSM}\\[5pt]
g_{\mathrm{DSM}}^\ast&=0\,,\\[5pt]
\lambda_{\mathrm{DSM}}^\ast&=\frac{8D}{c_DN_f}\frac{(4-D)(D-2)^2}{(3D-4)^2}\,.
\end{align}
The values of the couplings at the DSM fixed point depend only on the dimension $D$ and the fermion number $N_f$.
As we mentioned before, they are independent of the order of truncation within the LPA${^\prime}$ and the fixed point solution can be generalized to arbitrary $\lambda_{r,s}$.
\begin{align}
\lambda_{r,s}^{*\mathrm{DSM}}=\left(\frac{(4-D)(2-D)D}{c_DN_f(3D-4)}\right)^r\frac{4c_D N_fr!}{D(2r-D)}\delta_{s,0}\;,
\end{align}
for all $r,s\in\mathbb{N}$.
Further, the anomalous dimensions for the oder parameter fluctuations and the Dirac fermions characterizing the DSM are
\begin{align}
\eta_\phi^*&=4-D\,,\\[5pt]
\eta_\psi^*&=0\,.
\end{align}
%
%
An evaluation of the stability matrix at the DSM fixed point gives three leading eigenvalues
\begin{align}
	\theta_{1,\mathrm{DSM}}=D-3\,,\quad \theta_{2,\mathrm{DSM}}=\theta_{3,\mathrm{DSM}}=D-4\,.
\end{align}
Therefore, disregarding the flow of the mass term $\sim \overline{m}^2$, there is no relevant direction at the DSM fixed point as $\theta_i\leq 0$. Only the direction corresponding to the cubic coupling turns out to be marginal.
This implies that, while all other couplings flow to their respective DSM-fixed-point values, the cubic coupling freezes at different infrared values, depending on the initial conditions.
%

\subsubsection{Symmetry-broken regime} \label{sec:SSBflow}

After departing from the FIQCP in the symmetry-broken regime, the RG flow exhibits a window of scales where the couplings are dominated by the fermionic generalization of the Nambu-Goldstone fixed point, which is well-known from the purely bosonic O(N) models.
Therefore, it is characterized by the vanishing of all the U(1) breaking couplings $\Lambda_{i,j}=0$ for $j>0$ {and $\overline{h}^{2}=0$}.
Furthermore, we can define it formally in terms of the limit $\overline{\kappa}_\rho\to\infty$, which allows us to simplify the $\beta$ functions in the symmetry-broken regime
\begin{align}
\beta_{\Lambda_{2,0}}&=\Lambda_{2,0}\left(\frac{8v_D\Lambda_{2,0}}{D}+(D-4)\right)\;,\label{eq:NGSSB1}\\
\beta_{\Lambda_{3,0}}&=(2D-6)\Lambda_{3,0}+\frac{24v_D\Lambda_{2,0}}{D}\left(\Lambda_{3,0}-\Lambda_{2,0}^2\right)\;.
\label{eq:NGSSB2}
\end{align}

\begin{figure}[t!]
\includegraphics[width=.89\columnwidth]{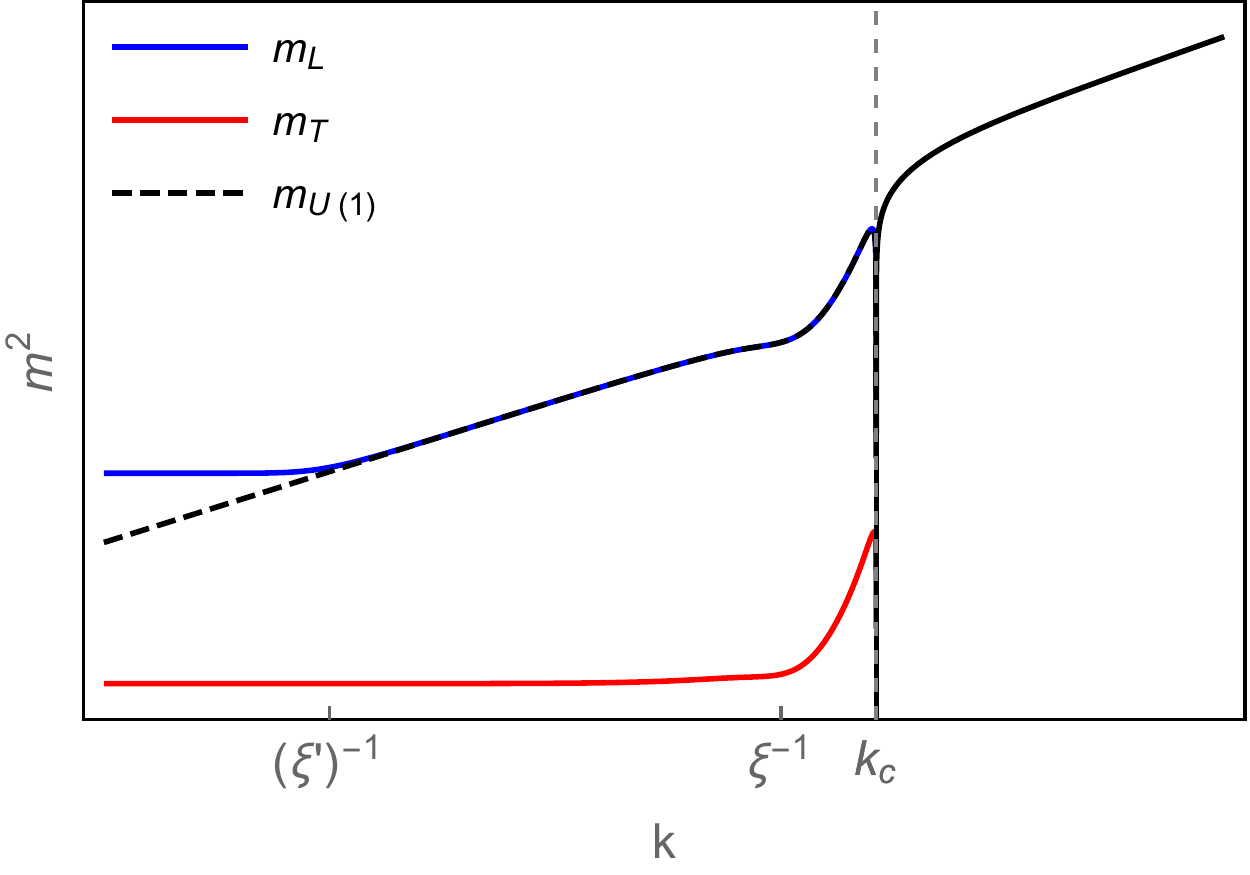}
\caption{\textbf{Flow of the dimensionful masses} $m^2_{L,T}$, cf. Eqs.~ \eqref{eq:massL} and \eqref{eq:massT}, in the symmetry broken regime. The red and blue curves correspond to the masses of one of the trajectories that escape the NG fixed point (i.e., when $g\neq0$.) and the black dashed curve corresponds to the flow of the purely $U(1)$ symmetric system with otherwise identical initial conditions. At the beginning of the flow all masses are identical and they eventually split when approaching the IR. In the IR both masses flow to non-zero values $m_L^2(k=0)\neq0$, $m_T^2(k=0)\neq0$.}
\label{fig:twomasses2}
\end{figure}

These $\beta$ functions admit a non-trivial fixed point solution for the scalar couplings, which we refer to as the Nambu-Goldstone (NG) fixed point, reading
\begin{align}
\Lambda_{2,0}^{\ast \mathrm{NG}}=\frac{D(4-D)}{8v_D},\ \Lambda_{3,0}^{\ast \mathrm{NG}}=3\left(\frac{D}{8v_D}\right)^2\frac{(4-D)^3}{6-D}\,.
\label{eq:NGSSB3}
\end{align}
Just as in the case of the DSM fixed points, the $\beta$ functions \eqref{eq:NGSSB2}, and therefore the fixed point values of the couplings, \eqref{eq:NGSSB3}, are independent of the order of the truncation and are valid for any LPA$n^\prime$. This can be traced back to the fact that for $\overline{h}^2\to0$ and $\overline{\kappa}_\rho\to\infty$, the $\beta$ functions for any given coupling of order $r$ depend only on the couplings of degrees less than $r$, i.e. there is some function depending on $r$, $f_r$, such that
\begin{align}
\beta_{\Lambda_{r,0}}=f_r(\{\Lambda_{j,0}\}),\quad j\leq r\;.
\end{align}
Moreover, it can be seen that the NG fixed point solution is independent of $N_f$.
The Yukawa coupling and the U(1) breaking coupling are both relevant directions at the NG fixed point.
Without them, i.e. in case of the O(2) model, the NG fixed point is fully attractive and completely dominates the infrared behavior of the model in the symmetry-broken phase.
The evaluation of the full stability matrix at the NG fixed point, taking into account perturbations in the direction induced by $\overline{h}^2$ and U(1)-breaking couplings $\Lambda_{i,j}^{\ast\mathrm{NG}}=0$ for $j>0$, shows that the cubic coupling not only is relevant at the NG fixed point, but it is the most relevant coupling in $D=3$, as
\begin{align}
[g]_{\mathrm{NG}}=\frac{6-D}{2}\;,\ [h^2]_{\mathrm{NG}}=4-D \;,\ [\kappa_{\rho}]_\mathrm{NG}=2-D\;.
\end{align}
The fact that the cubic coupling changes from being irrelevant at the FIQCP to relevant at the NG fixed point implies that eventually, in the deep infrared, the flow is driven away from the NG fixed point when the cubic coupling does not exactly vanish.
This means that $g$ will start to grow. As a result the transversal mass, which is zero for $g=0$, also increases and thus provides a second mass scale besides the longitudinal mass (cf.~Eq.~\eqref{eq:massL}). Consequently, a second correlation length can be defined as $\bar m_T^2(k=\xi'^{-1})\approx 1$, cf. Ref.~\onlinecite{PhysRevLett.115.200601}.
We show the evolution of the longitudinal and transversal mass in Fig.~\ref{fig:twomasses2}.
For $k>k_c$, the longitudinal and transversal mass are identical and both flow to zero at the transition to the Kekul\'e phase. On the symmetry-broken side of the transition $k<k_c$, they split and, after an initial increase, fluctuations lead to a decrease of the masses towards the infrared. Longitudinal and transversal fluctuations cease to contribute to the flow when $k^2\lesssim m_i^2$ (i.e. the dimensionless $\overline{m}_i^2\gtrsim 1$). This happens first for $\overline{m}_L^2$ when the flow transverses from the FIQCP to the NG FP. At this point, the ``common'' characteristic correlation length $\xi$ that also appears when a continuous $U(1)$ symmetry is broken can be defined $\overline{m}_L^2(k\sim \xi^{-1})=1$. Below this scale, $k<\xi^{-1}$, the (pseudo) Goldstone modes drive the reduction of $m_L^2$. Approximately at the same time, $m_T^2$ stops running because the cubic coupling and the minimum reach their infrared values (cf. Eq.~\eqref{eq:massT}). As we are close to the NG FP, the infrared value of the cubic coupling is small, but nonzero. Therefore towards the infrared when $k\rightarrow 0$, the transversal mass becomes $m_T^2>k^2$. At this point, which defines the second length scale $\overline{m}_T^2(k\sim \xi^{-1})=1$, the flow leaves the NG FP and the longitudinal mass also reaches a finite value. This is different to the flow of a $U(1)$-symmetric system, where the Goldstone modes would eventually drive the longitudinal mass to zero.

\subsection{Second correlation length exponent}

Having established the global RG flow of the model, we turn again to its critical properties.
In Tab.~\ref{tab:critexp}, we show the two largest critical exponents for several choices of $N_f$. Here, $\nu=\theta_1^{-1}$ corresponds to the inverse correlation length exponent of the $\mathbb{Z}_3$~order. $\theta_2$ is the second largest exponent and is related to the scaling of the U(1) breaking cubic coupling. It determines the stability of the FIQCP and dominates the corrections to scaling~\cite{PhysRevB.96.115132}.
Further, the appearance of two length scales in the symmetry-broken regime, $\xi$ and $\xi^\prime$, is related to the presence of two correlation length exponents $\nu$ and $\nu^\prime$.
Despite the exclusive appearance of the second length scale in the symmetry-broken regime, the ratio of their correlation length exponents $\nu^\prime/\nu$ is uniquely determined by a scaling law employing properties of the FIQCP alone~\cite{PhysRevLett.115.200601,PhysRevB.96.195162}
\begin{align}
	\frac{\nu^\prime}{\nu}=1-\frac{\theta_2}{2}\,.
\end{align}
This can be understood in terms of the dangerously irrelevant coupling $g$. Only if $g$ is non-zero, the transversal mass, which is responsible for the second scale, will be finite away from the critical point. Therefore the scaling of $m_T^2$ must be related to scaling of $g$ at the critical point. This scaling, in turn, is given by $\theta_2$.
We list our estimates for $\nu^\prime$ also in Tab.~\ref{tab:critexp}, whenever we find that the FIQCP is stable in $D=2+1$. $\nu^\prime$ shows only very small deviations from $\nu$ due to the smallness of $|\theta_2|\ll 1$ which we deem unlikely to be observed either in an experimental setup or in lattice QMC simulations.
Instead, as stated earlier~\cite{PhysRevB.96.115132}, we expect that large corrections to scaling will have to be considered in corresponding lattice QMC simulations.
%

\begin{table}[t!]
\caption{\label{tab:critexp} \textbf{Correlation length exponents:} Numerical values for the largest two critical exponents and the second correlation length exponent for different $N_f$ in $D=2+1$ in LPA$12^\prime$. The exponents are given by $\nu=\theta_1^{-1}$ and $\nu^\prime=\nu(1-\theta_2/2)$. The exponent deciding over stability $\theta_2$ is shown in boldface.}
\begin{tabular*}{\linewidth}{@{\extracolsep{\fill} } l l l l }
\hline\hline
  $N_f$ & $\nu$ & $\theta_2$ & $\nu^\prime$\\ \hline
1 & 1.195 & {\bf +0.167} & -\\
2 & 1.157 & {\bf -0.0031} & 1.159\\
3 & 1.109 & {\bf -0.0235} & 1.122\\
4 & 1.082 & {\bf -0.0263} & 1.096\\
5 & 1.066 & {\bf -0.0255} & 1.080\\
$\infty$ & 1 & {\bf 0} & 1\\
 \hline\hline
\end{tabular*}
\end{table}

\section{Conclusion}\label{sec:conc}

In this work, we have provided a thorough study of the renormalization group flow near the fluctuation-induced quantum phase transition to the Kekul\'e VBS state in Dirac semi-metals as it appears for fermions on the two-dimensional honeycomb lattice.
This transition is characterized by the condensation of the Kekul\'e order parameter which reduces the chiral U(1) symmetry of the Dirac system to a discrete $\mathbb{Z}_3$ symmetry, resulting in a series of unconventional properties at and close to the quantum transition.
Firstly, the fact that the transition is continuous and not discontinuous is an effect driven by strong (fermion) fluctuations. Secondly, there is an emergent U(1) symmetry at the quantum critical point and, thirdly, a second length scale appears in the symmetry broken phase due to the breaking of the discrete symmetry.
We noted that these properties are shared with the scenario of the deconfined QCPs.

We have investigated the semimetal-to-Kekul\'e quantum transition in terms of an appropriate Gross-Neveu-Yukawa model with the help of the non-perturbative functional renormalization group.
While building on our previous work, we have extended it in various directions.
First, we have established the FIQCP scenario for the full range of dimensions between $1+1 \leq D \leq 2+1$ for small numbers of fermion flavors $N_f\leq 2$.
We have found that for every $N_f>0$, there is a critical dimension $2<D<4$ above which the fermion-enhanced O(2) fixed point becomes stable, giving rise to a second order transition.
This is in stark contrast to the model where $N_f=0$, corresponding to the field-theoretical formulation of the three-state Potts model, where only close to $D=1+1$ does a stable O(2)-breaking fixed point appear.
Our findings fit nicely to the observation that fermions in Dirac systems tend to support symmetry enhancement~\cite{Ponte:2012ru,Grover280,Roy:2015zna,Li:2017dkj,PhysRevB.97.041117}.

Furthermore, we have discussed that, due to the discrete symmetry breaking, there are no Goldstone modes.
Instead, in the symmetry-broken regime, two finite masses appear - the longitudinal and transversal mass.
Importantly, the scaling of the transversal mass depends on the scaling of the cubic term and defines a second length scale, which diverges at the QCP.
We provided a comprehensive analysis of the fixed-point structure in the symmetric as well as in the symmetry broken regime.
By solving the flow equations in both regimes, we studied  the complete behavior of the RG flow beyond a pure fixed point analysis.
While the Dirac semimetal fixed point dominates the long-range behavior of the system in the semimetallic phase, the symmetry-broken regime shows a more subtle behavior:
here, the Nambu-Goldstone fixed point dominates the system on intermediate length scales.
Eventually, when the cubic coupling is finite, the RG flow leaves the NG regime giving rise to a sizable transversal mass and the concomitant second length scale.
We also calculated improved estimates for the correlation length exponent of the second length scale and show that it is very close to the order parameter correlation length exponent.
We therefore expect that it will be very challenging to observe this behavior in numerical simulations.
In summary, we provided a unified picture of the system close to the fermion-induced QCP in the symmetric as well as in the symmetry-broken phases.\bigskip
%

\paragraph*{Acknowledgments.}
The authors are grateful to Fabian Rennecke and Riccardo Ben Ali Zinati for discussions.
MMS was supported by the DFG through the Collaborative
Research Center SFB1238, TP C04. ET was supported by the DFG through the Leibniz Prize of A. Rosch, RO2265/5-1. Work at BNL is supported by the U.S. Department of Energy (DOE), Division of Condensed Matter Physics and Materials Science under Contract No. DE-AC02-98CH10886. IFH is supported by the NSERC of Canada.

\appendix

\section{Mass terms and effective potential}\label{app:flow}

The flow equations are functions of the second and third derivatives of the potential. These are denoted by $u_{ij}=\frac{\partial^2 u}{\partial\phi_i\partial\phi_j}$ and analogously for $u_{ijk}$. We moreover denote derivatives with respect to the invariants Eqs.~\eqref{eq:invariants} and \eqref{eq:newtau} as $u^{(m,n)}=\frac{\partial^{m+n} u}{\partial\rho^m\partial\tau^n}$. With these conventions, the general expression for the masses in terms of the invariants $\rho$ and $\tau$ is
\begin{widetext}
\begin{align}
m_{L,T}^2&=\rho  u^{(2,0)}+\frac{9}{2} \sqrt{\rho }u^{(0,1)}+u^{(1,0)}+ \tau\left(2u^{(0,2)}+3u^{(1,1)}\right)\pm\nonumber\\
&\left[\left(\rho u^{(2,0)}-\frac{9}{2}\sqrt{\rho}u^{(0,1)}\right)^2+3\tau\left(2u^{(0,1)}+3 (\tau -2 \rho ^{3/2}) u^{(0,2)}+2 \rho u^{(1,1)}\right) u^{(2,0)}+ \frac{9\tau}{\sqrt{\rho}}(u^{(0,1)})^2\right.\nonumber\\
&\left.+\frac{9\tau\left(\left(3 (6 \rho ^{3/2}+\tau) u^{(0,2)}+10 \rho  u^{(1,1)}\right) u^{(0,1)}+2 \rho   \left(9 \sqrt{\rho } \tau  (u^{(0,2)})^2+6 \tau  u^{(1,1)} u^{(0,2)}+4 \rho  (u^{(1,1)})^2\right)\right)}{2\sqrt{\rho }}\right]^{1/2}\;.\label{eq:massesfull}
\end{align}
\end{widetext}
Moreover, for the choice of minimum described in the main text ($\phi_1^{\min} = - \sqrt{2\rho},\phi_2^{\min} =
0$), the terms appearing in the flow equations are given by the expressions
\begin{align}
u_{11}&=2\rho u^{(2,0)}+u^{(1,0)}\;,\\
u_{12}&=u_{21}=0\;,\\
u_{22}&=9\sqrt{\rho}u^{(0,1)}+u^{(1,0)}\;,
\end{align}
and $\omega_\psi=2 h^2\rho$. Further,
\begin{align}
u_{111}&=-\sqrt{2\rho}\left( 3u^{(2,0)}+2\rho u^{(3,0)}\right)\;,\\
u_{112}&=u_{121}=u_{211}=u_{222}=0\,,\\
u_{221}&=u_{212}=u_{122}=\frac{9 u^{(0,1)}+18\rho u^{(1,1)}+2\sqrt{\rho}u^{(2,0)}}{-\sqrt{2}}\;.\notag
\end{align}
This completes the set of required definitions.

\section{Threshold functions}\label{app:flow2}

Here we present the threshold functions that appear in the flow equations and the anomalous dimensions.
In the following we restrict to a choice of cutoff $R_{\Phi,k}(q)$ that allows explicit analytic evaluation of the integrals involved and has favorable convergence properties~\cite{Litim:2000ci,Litim:2001up,Litim:2001fd,Litim:2002cf,PAWLOWSKI2017165}:
\begin{align}
R_{\phi,k}(q)&=Z_{\phi,k}(k^2-q^2)\theta(k^2-q^2)\,,\\
qR_{\psi,k}(q)&=iZ_{\psi,k}\slashed{q}(k-q)\theta(k^2-q^2)\,.
\end{align}
The threshold functions in the effective potential read
\begin{align}
l_B(\omega)&=\frac{2}{D}\left(1-\frac{\eta_\phi}{D+2}\right)\frac{1}{1+\omega}\,,\\
l_F(\omega)&=\frac{2}{D}\left(1-\frac{\eta_\psi}{D+1}\right)\frac{1}{1+\omega}\,.
\end{align}
In the anomalous dimensions the threshold functions are
\begin{align}
m_{2}^F&=\frac{1}{(1+\omega_\psi)^4}\,,\\
m_{4}^F&=\frac{1}{(1+\omega_\psi)^4} + \frac{1-\eta_\psi}{D-2}\frac{1}{(1+\omega_\psi)^3}\notag \\
&\quad- \left( \frac{1-\eta_\psi}{2D-4} +\frac{1}{4} \right)\frac{1}{(1+\omega_\psi)^2}\,.
\end{align}
For any of the three equivalent minima of the order parameter potential, the remaining threshold functions are somewhat more involved and read

\begin{widetext}

\begin{align}
m_{(4)R_{1/2}}^B(v_1,v_2)&=\left( (1+u_{22/11})\frac{(1+u_{22/11})v_1-u_{12}v_2}{((1+u_{11})(1+u_{22})-u_{12}^2)^2} -u_{12}\frac{(1+u_{11/22})v_2-u_{12}v_1}{((1+u_{11})(1+u_{22})-u_{12}^2)^2}\right)^2\,,\\
m_{(22)R_1R_2}^B(v_1,v_2,v_3)&= (1+u_{11})(1+u_{22})\frac{((1+u_{11})v_1-u_{12}v_2)((1+u_{22})v_1-u_{12}v_3)}{((1+u_{11})(1+u_{22})-u_{12}^2)^4} \nonumber \\
&\quad+u_{12}^2\frac{((1+u_{22})v_2-u_{12}v_1)((1+u_{11})v_3-u_{12}v_1)}{((1+u_{11})(1+u_{22})-u_{12}^2)^4} \nonumber \\
&\quad-u_{12}(1+u_{22})\frac{((1+u_{22})v_2-u_{12}v_1)((1+u_{22})v_1-u_{12}v_3)}{((1+u_{11})(1+u_{22})-u_{12}^2)^4} \nonumber \\
&\quad-u_{12}(1+u_{11})\frac{((1+u_{11})v_1-u_{12}v_2)((1+u_{11})v_3-u_{12}v_1)}{((1+u_{11})(1+u_{22})-u_{12}^2)^4}\,, \\
m_{(12)R_{1/2}}^{FB}&=\left( 1-\frac{\eta_\phi}{D+1} \right)\frac{1}{\Big((1+u_{11})(1+u_{22})-u_{12}^2\Big)^2(1+\omega_\psi)} \Big( (1+u_{22/11})^2+ u_{12}^2 \Big)\,,
\end{align}
\begin{align}
l_{nm}^{FR_{1/2}}=&\frac{2}{D}\left[ m\left( 1-\frac{\eta_\phi}{D+2} \right)\frac{(1+u_{22/11})^2+u_{12}^2}{\Big((1+u_{11})(1+u_{22})-u_{12}^2\Big)(1+u_{22/11})} + n\left( 1-\frac{\eta_\psi}{D+1} \right)\frac{1}{1+\omega_\psi} \right] \nonumber \\
& \times \frac{(1+u_{22/11})^m}{\Big((1+u_{11})(1+u_{22})-u_{12}^2\Big)^m(1+\omega_\psi)^n}\,, \\
l_{12}^{FA}=&\frac{2}{D}\left[2\left( 1-\frac{\eta_\phi}{D+2} \right)\frac{(1+u_{11})+(1+u_{22})}{(1+u_{11})(1+u_{22})-u_{12}^2} + \left( 1-\frac{\eta_\psi}{D+1} \right)\frac{1}{1+\omega_\psi}\right] \frac{u_{12}^2}{\Big((1+u_{11})(1+u_{22})-u_{12}^2\Big)^2(1+\omega_\psi)}\,, \\
l_{111}^{FR_{1/2}A}=&\frac{2}{D}\left[\left( 1-\frac{\eta_\phi}{D+2} \right)\frac{2(1+u_{22/11})^2+(1+u_{11})(1+u_{22})+u_{12}^2}{(1+u_{11})(1+u_{22})-u_{12}^2} + \left( 1-\frac{\eta_\psi}{D+1} \right)\frac{1+u_{22/11}}{1+\omega_\psi}\right] \nonumber \\
&\times \frac{u_{12}}{\Big((1+u_{11})(1+u_{22})-u_{12}^2\Big)^2(1+\omega_\psi)}\,, \\
l_{111}^{FR_1R_2}=&\frac{2}{D}\left[\left( 1-\frac{\eta_\phi}{D+2} \right)(2+u_{11}+u_{22})\frac{(1+u_{11})(1+u_{22})+u_{12}^2}{(1+u_{11})(1+u_{22})-u_{12}^2} + \left( 1-\frac{\eta_\psi}{D+1} \right)\frac{(1+u_{11})(1+u_{22})}{1+\omega_\psi}\right] \nonumber \\
&\times \frac{1}{\Big((1+u_{11})(1+u_{22})-u_{12}^2\Big)^2(1+\omega_\psi)}\,.
\end{align}
Let us note that these threshold functions simplify for our choice of the minimum as $u_{12}=0$ in this case.
\end{widetext}

\section{Remark on the Potts model}\label{app:potts}

In two dimensions, there are exact results for the three-state Potts models' critical exponents. More specifically, it was found that there is a critical Potts and a tricritical fixed point (TCP).
The Potts FP comes with correlation length exponent and anomalous dimension\cite{RevModPhys.54.235}
\begin{align}
\mathrm{Potts}:&\ \nu_{2,\mathrm{ex}}=\frac{5}{6}\approx 0.83\,,\ \eta_{2,\mathrm{ex}}=\frac{4}{15}\approx 0.27\,.
\end{align}
The critical and tricritical fixed points can be continued above two dimensions where they change their coordinates.
It was found\cite{} that they collide and disappear to the complex plane for some $d_c>2$.
Then no stable fixed point exists in the system and the transition from the symmetric to the $\mathbb{Z}_3$ ordered phase is discontinuous.
Numerical results in $d=3$ for the three-state Potts model suggest that $d_c<3$.

Here, we explore a simple approach to the two-dimensional case by a finite expansion in the LPA${}^\prime$ and note that this can only provide qualitative results on the Potts fixed point.
As has been shown in earlier work for scalar~\cite{PhysRevLett.75.378,PhysRevB.64.054513,Codello:2012sc,PhysRevLett.110.141601,Borchardt:2016kco} and scalar-fermion models~\cite{Janssen:2014gea}, the FRG can also be used to describe critical behavior below three dimensions.
For a quantitive estimate, it will be required to work with higher expansions or use methods beyond a finite expansion, as towards two dimensions more and more couplings become canonically relevant.

Employing an LPA8${}^\prime$ expansion in the symmetric regime, we can give first estimates on the limit of the purely bosonic Potts model.
In fact, we find both the critical Potts fixed point and one tricritial fixed point, as expected.
In $d=2$ the Potts fixed has the critical exponents
\begin{align}
 \mathrm{Potts}:&\ \nu_{2,\mathrm{LPA8}^\prime} \approx 0.82\,,\quad \eta_{2,\mathrm{LPA8}^\prime} \approx 0.22\,,
\end{align}
which already compares well to the exact results.
Also, we can continue the fixed point search in higher dimensions towards $d=3$ and beyond to exhibit the fixed-point collision. Within LPA8${}^\prime$, we find that the Potts FP and the tricritical FP indeed collide at a critical dimension of $d_{c,\mathrm{LPA8}^\prime}\approx 3.2$ which seems too large in comparison with numerical results. We observe that higher orders in the LPA${}^\prime$ seem to push the critical dimension below three.
However, we note that, in contrast to the FIQCP,  going to higher orders in the LPA${}^\prime$ still leads to corrections and we do not yet see convergence.
Therefore, to settle the convergence of the critical exponents and the critical dimension for the Pott's fixed point with the fRG,  a more thorough study will be required, which is beyond the scope of this work. As one alternative, we suggest the pseudo-spectral methods as developed in Refs.~\onlinecite{Borchardt:2015rxa,Borchardt:2016pif,Borchardt:2016kco} for FRG applications.

\bibliography{flowSSB}

\begin{thebibliography}{85}%
\makeatletter
\providecommand \@ifxundefined [1]{%
 \@ifx{#1\undefined}
}%
\providecommand \@ifnum [1]{%
 \ifnum #1\expandafter \@firstoftwo
 \else \expandafter \@secondoftwo
 \fi
}%
\providecommand \@ifx [1]{%
 \ifx #1\expandafter \@firstoftwo
 \else \expandafter \@secondoftwo
 \fi
}%
\providecommand \natexlab [1]{#1}%
\providecommand \enquote  [1]{``#1''}%
\providecommand \bibnamefont  [1]{#1}%
\providecommand \bibfnamefont [1]{#1}%
\providecommand \citenamefont [1]{#1}%
\providecommand \href@noop [0]{\@secondoftwo}%
\providecommand \href [0]{\begingroup \@sanitize@url \@href}%
\providecommand \@href[1]{\@@startlink{#1}\@@href}%
\providecommand \@@href[1]{\endgroup#1\@@endlink}%
\providecommand \@sanitize@url [0]{\catcode `\\12\catcode `\$12\catcode
  `\&12\catcode `\#12\catcode `\^12\catcode `\_12\catcode `\%12\relax}%
\providecommand \@@startlink[1]{}%
\providecommand \@@endlink[0]{}%
\providecommand \url  [0]{\begingroup\@sanitize@url \@url }%
\providecommand \@url [1]{\endgroup\@href {#1}{\urlprefix }}%
\providecommand \urlprefix  [0]{URL }%
\providecommand \Eprint [0]{\href }%
\providecommand \doibase [0]{http://dx.doi.org/}%
\providecommand \selectlanguage [0]{\@gobble}%
\providecommand \bibinfo  [0]{\@secondoftwo}%
\providecommand \bibfield  [0]{\@secondoftwo}%
\providecommand \translation [1]{[#1]}%
\providecommand \BibitemOpen [0]{}%
\providecommand \bibitemStop [0]{}%
\providecommand \bibitemNoStop [0]{.\EOS\space}%
\providecommand \EOS [0]{\spacefactor3000\relax}%
\providecommand \BibitemShut  [1]{\csname bibitem#1\endcsname}%
\let\auto@bib@innerbib\@empty
\bibitem [{\citenamefont {Landau}\ and\ \citenamefont
  {Lifshitz}(1980)}]{Landau:1980mil}%
  \BibitemOpen
  \bibfield  {author} {\bibinfo {author} {\bibfnamefont {L.~D.}\ \bibnamefont
  {Landau}}\ and\ \bibinfo {author} {\bibfnamefont {E.~M.}\ \bibnamefont
  {Lifshitz}},\ }\href@noop {} {\emph {\bibinfo {title} {{Statistical Physics,
  Part 1}}}},\ \bibinfo {series} {Course of Theoretical Physics}, Vol.~\bibinfo
  {volume} {5}\ (\bibinfo  {publisher} {Butterworth-Heinemann},\ \bibinfo
  {address} {Oxford},\ \bibinfo {year} {1980})\BibitemShut {NoStop}%
\bibitem [{\citenamefont {Wegner}\ and\ \citenamefont
  {Houghton}(1973)}]{Wegner:1972ih}%
  \BibitemOpen
  \bibfield  {author} {\bibinfo {author} {\bibfnamefont {F.~J.}\ \bibnamefont
  {Wegner}}\ and\ \bibinfo {author} {\bibfnamefont {A.}~\bibnamefont
  {Houghton}},\ }\bibfield  {title} {{\color{Gray}\small \bibinfo {title}
  {{Renormalization group equation for critical phenomena}},\ }}\href {\doibase
  10.1103/PhysRevA.8.401} {\bibfield  {journal} {\bibinfo  {journal} {Phys.
  Rev.}\ }\textbf {\bibinfo {volume} {A8}},\ \bibinfo {pages} {401} (\bibinfo
  {year} {1973})}\BibitemShut {NoStop}%
\bibitem [{\citenamefont {Wilson}\ and\ \citenamefont
  {Kogut}(1974)}]{Wilson:1973jj}%
  \BibitemOpen
  \bibfield  {author} {\bibinfo {author} {\bibfnamefont {K.~G.}\ \bibnamefont
  {Wilson}}\ and\ \bibinfo {author} {\bibfnamefont {J.~B.}\ \bibnamefont
  {Kogut}},\ }\bibfield  {title} {{\color{Gray}\small \bibinfo {title} {{The
  Renormalization group and the epsilon expansion}},\ }}\href {\doibase
  10.1016/0370-1573(74)90023-4} {\bibfield  {journal} {\bibinfo  {journal}
  {Phys. Rept.}\ }\textbf {\bibinfo {volume} {12}},\ \bibinfo {pages} {75}
  (\bibinfo {year} {1974})}\BibitemShut {NoStop}%
\bibitem [{\citenamefont {Senthil}\ \emph
  {et~al.}(2004{\natexlab{a}})\citenamefont {Senthil}, \citenamefont {Balents},
  \citenamefont {Sachdev}, \citenamefont {Vishwanath},\ and\ \citenamefont
  {Fisher}}]{PhysRevB.70.144407}%
  \BibitemOpen
  \bibfield  {author} {\bibinfo {author} {\bibfnamefont {T.}~\bibnamefont
  {Senthil}}, \bibinfo {author} {\bibfnamefont {L.}~\bibnamefont {Balents}},
  \bibinfo {author} {\bibfnamefont {S.}~\bibnamefont {Sachdev}}, \bibinfo
  {author} {\bibfnamefont {A.}~\bibnamefont {Vishwanath}}, \ and\ \bibinfo
  {author} {\bibfnamefont {M.~P.~A.}\ \bibnamefont {Fisher}},\ }\bibfield
  {title} {{\color{Gray}\small \bibinfo {title} {{Quantum criticality beyond
  the Landau-Ginzburg-Wilson paradigm}},\ }}\href {\doibase
  10.1103/PhysRevB.70.144407} {\bibfield  {journal} {\bibinfo  {journal} {Phys.
  Rev. B}\ }\textbf {\bibinfo {volume} {70}},\ \bibinfo {pages} {144407}
  (\bibinfo {year} {2004}{\natexlab{a}})}\BibitemShut {NoStop}%
\bibitem [{\citenamefont {Senthil}\ \emph
  {et~al.}(2004{\natexlab{b}})\citenamefont {Senthil}, \citenamefont
  {Vishwanath}, \citenamefont {Balents}, \citenamefont {Sachdev},\ and\
  \citenamefont {Fisher}}]{Senthil1490}%
  \BibitemOpen
  \bibfield  {author} {\bibinfo {author} {\bibfnamefont {T.}~\bibnamefont
  {Senthil}}, \bibinfo {author} {\bibfnamefont {A.}~\bibnamefont {Vishwanath}},
  \bibinfo {author} {\bibfnamefont {L.}~\bibnamefont {Balents}}, \bibinfo
  {author} {\bibfnamefont {S.}~\bibnamefont {Sachdev}}, \ and\ \bibinfo
  {author} {\bibfnamefont {M.~P.~A.}\ \bibnamefont {Fisher}},\ }\bibfield
  {title} {{\color{Gray}\small \bibinfo {title} {{Deconfined Quantum Critical
  Points}},\ }}\href {\doibase 10.1126/science.1091806} {\bibfield  {journal}
  {\bibinfo  {journal} {Science}\ }\textbf {\bibinfo {volume} {303}},\ \bibinfo
  {pages} {1490} (\bibinfo {year} {2004}{\natexlab{b}})}\BibitemShut {NoStop}%
\bibitem [{\citenamefont {Senthil}\ \emph {et~al.}(2005)\citenamefont
  {Senthil}, \citenamefont {Balents}, \citenamefont {Sachdev}, \citenamefont
  {Vishwanath},\ and\ \citenamefont {Fisher}}]{doi:10.1143/JPSJS.74S.1}%
  \BibitemOpen
  \bibfield  {author} {\bibinfo {author} {\bibfnamefont {T.}~\bibnamefont
  {Senthil}}, \bibinfo {author} {\bibfnamefont {L.}~\bibnamefont {Balents}},
  \bibinfo {author} {\bibfnamefont {S.}~\bibnamefont {Sachdev}}, \bibinfo
  {author} {\bibfnamefont {A.}~\bibnamefont {Vishwanath}}, \ and\ \bibinfo
  {author} {\bibfnamefont {M.~P.~A.}\ \bibnamefont {Fisher}},\ }\bibfield
  {title} {{\color{Gray}\small \bibinfo {title} {{Deconfined Criticality
  Critically Defined}},\ }}\href {\doibase 10.1143/JPSJS.74S.1} {\bibfield
  {journal} {\bibinfo  {journal} {Journal of the Physical Society of Japan}\
  }\textbf {\bibinfo {volume} {74}},\ \bibinfo {pages} {1} (\bibinfo {year}
  {2005})},\ \Eprint {http://arxiv.org/abs/https://doi.org/10.1143/JPSJS.74S.1}
  {https://doi.org/10.1143/JPSJS.74S.1} \BibitemShut {NoStop}%
\bibitem [{\citenamefont {{Li}}\ \emph {et~al.}(2017)\citenamefont {{Li}},
  \citenamefont {{Jiang}}, \citenamefont {{Jian}},\ and\ \citenamefont
  {{Yao}}}]{2015arXiv151207908L}%
  \BibitemOpen
  \bibfield  {author} {\bibinfo {author} {\bibfnamefont {Z.-X.}\ \bibnamefont
  {{Li}}}, \bibinfo {author} {\bibfnamefont {Y.-F.}\ \bibnamefont {{Jiang}}},
  \bibinfo {author} {\bibfnamefont {S.-K.}\ \bibnamefont {{Jian}}}, \ and\
  \bibinfo {author} {\bibfnamefont {H.}~\bibnamefont {{Yao}}},\ }\bibfield
  {title} {{\color{Gray}\small \bibinfo {title} {{Fermion-induced quantum
  critical points}},\ }}\href
  {https://www.nature.com/articles/s41467-017-00167-6} {\bibfield  {journal}
  {\bibinfo  {journal} {Nature Communications}\ }\textbf {\bibinfo {volume}
  {8}},\ \bibinfo {pages} {314} (\bibinfo {year} {2017})}\BibitemShut {NoStop}%
\bibitem [{\citenamefont {Scherer}\ and\ \citenamefont
  {Herbut}(2016)}]{PhysRevB.94.205136}%
  \BibitemOpen
  \bibfield  {author} {\bibinfo {author} {\bibfnamefont {M.~M.}\ \bibnamefont
  {Scherer}}\ and\ \bibinfo {author} {\bibfnamefont {I.~F.}\ \bibnamefont
  {Herbut}},\ }\bibfield  {title} {{\color{Gray}\small \bibinfo {title}
  {{Gauge-field-assisted Kekul\'e quantum criticality}},\ }}\href {\doibase
  10.1103/PhysRevB.94.205136} {\bibfield  {journal} {\bibinfo  {journal} {Phys.
  Rev. B}\ }\textbf {\bibinfo {volume} {94}},\ \bibinfo {pages} {205136}
  (\bibinfo {year} {2016})}\BibitemShut {NoStop}%
\bibitem [{\citenamefont {Jian}\ and\ \citenamefont
  {Yao}(2017{\natexlab{a}})}]{PhysRevB.96.195162}%
  \BibitemOpen
  \bibfield  {author} {\bibinfo {author} {\bibfnamefont {S.-K.}\ \bibnamefont
  {Jian}}\ and\ \bibinfo {author} {\bibfnamefont {H.}~\bibnamefont {Yao}},\
  }\bibfield  {title} {{\color{Gray}\small \bibinfo {title} {{Fermion-induced
  quantum critical points in two-dimensional Dirac semimetals}},\ }}\href
  {\doibase 10.1103/PhysRevB.96.195162} {\bibfield  {journal} {\bibinfo
  {journal} {Phys. Rev. B}\ }\textbf {\bibinfo {volume} {96}},\ \bibinfo
  {pages} {195162} (\bibinfo {year} {2017}{\natexlab{a}})}\BibitemShut
  {NoStop}%
\bibitem [{\citenamefont {Jian}\ and\ \citenamefont
  {Yao}(2017{\natexlab{b}})}]{PhysRevB.96.155112}%
  \BibitemOpen
  \bibfield  {author} {\bibinfo {author} {\bibfnamefont {S.-K.}\ \bibnamefont
  {Jian}}\ and\ \bibinfo {author} {\bibfnamefont {H.}~\bibnamefont {Yao}},\
  }\bibfield  {title} {{\color{Gray}\small \bibinfo {title} {{Fermion-induced
  quantum critical points in three-dimensional Weyl semimetals}},\ }}\href
  {\doibase 10.1103/PhysRevB.96.155112} {\bibfield  {journal} {\bibinfo
  {journal} {Phys. Rev. B}\ }\textbf {\bibinfo {volume} {96}},\ \bibinfo
  {pages} {155112} (\bibinfo {year} {2017}{\natexlab{b}})}\BibitemShut
  {NoStop}%
\bibitem [{\citenamefont {Classen}\ \emph {et~al.}(2017)\citenamefont
  {Classen}, \citenamefont {Herbut},\ and\ \citenamefont
  {Scherer}}]{PhysRevB.96.115132}%
  \BibitemOpen
  \bibfield  {author} {\bibinfo {author} {\bibfnamefont {L.}~\bibnamefont
  {Classen}}, \bibinfo {author} {\bibfnamefont {I.~F.}\ \bibnamefont {Herbut}},
  \ and\ \bibinfo {author} {\bibfnamefont {M.~M.}\ \bibnamefont {Scherer}},\
  }\bibfield  {title} {{\color{Gray}\small \bibinfo {title}
  {{Fluctuation-induced continuous transition and quantum criticality in Dirac
  semimetals}},\ }}\href {\doibase 10.1103/PhysRevB.96.115132} {\bibfield
  {journal} {\bibinfo  {journal} {Phys. Rev. B}\ }\textbf {\bibinfo {volume}
  {96}},\ \bibinfo {pages} {115132} (\bibinfo {year} {2017})}\BibitemShut
  {NoStop}%
\bibitem [{\citenamefont {Oshikawa}(2000)}]{PhysRevB.61.3430}%
  \BibitemOpen
  \bibfield  {author} {\bibinfo {author} {\bibfnamefont {M.}~\bibnamefont
  {Oshikawa}},\ }\bibfield  {title} {{\color{Gray}\small \bibinfo {title}
  {{Ordered phase and scaling in ${Z}_{n}$ models and the three-state
  antiferromagnetic Potts model in three dimensions}},\ }}\href {\doibase
  10.1103/PhysRevB.61.3430} {\bibfield  {journal} {\bibinfo  {journal} {Phys.
  Rev. B}\ }\textbf {\bibinfo {volume} {61}},\ \bibinfo {pages} {3430}
  (\bibinfo {year} {2000})}\BibitemShut {NoStop}%
\bibitem [{\citenamefont {Okubo}\ \emph {et~al.}(2015)\citenamefont {Okubo},
  \citenamefont {Oshikawa}, \citenamefont {Watanabe},\ and\ \citenamefont
  {Kawashima}}]{PhysRevB.91.174417}%
  \BibitemOpen
  \bibfield  {author} {\bibinfo {author} {\bibfnamefont {T.}~\bibnamefont
  {Okubo}}, \bibinfo {author} {\bibfnamefont {K.}~\bibnamefont {Oshikawa}},
  \bibinfo {author} {\bibfnamefont {H.}~\bibnamefont {Watanabe}}, \ and\
  \bibinfo {author} {\bibfnamefont {N.}~\bibnamefont {Kawashima}},\ }\bibfield
  {title} {{\color{Gray}\small \bibinfo {title} {{Scaling relation for
  dangerously irrelevant symmetry-breaking fields}},\ }}\href {\doibase
  10.1103/PhysRevB.91.174417} {\bibfield  {journal} {\bibinfo  {journal} {Phys.
  Rev. B}\ }\textbf {\bibinfo {volume} {91}},\ \bibinfo {pages} {174417}
  (\bibinfo {year} {2015})}\BibitemShut {NoStop}%
\bibitem [{\citenamefont {L\'eonard}\ and\ \citenamefont
  {Delamotte}(2015)}]{PhysRevLett.115.200601}%
  \BibitemOpen
  \bibfield  {author} {\bibinfo {author} {\bibfnamefont {F.}~\bibnamefont
  {L\'eonard}}\ and\ \bibinfo {author} {\bibfnamefont {B.}~\bibnamefont
  {Delamotte}},\ }\bibfield  {title} {{\color{Gray}\small \bibinfo {title}
  {{Critical Exponents Can Be Different on the Two Sides of a Transition: A
  Generic Mechanism}},\ }}\href {\doibase 10.1103/PhysRevLett.115.200601}
  {\bibfield  {journal} {\bibinfo  {journal} {Phys. Rev. Lett.}\ }\textbf
  {\bibinfo {volume} {115}},\ \bibinfo {pages} {200601} (\bibinfo {year}
  {2015})}\BibitemShut {NoStop}%
\bibitem [{\citenamefont {Shao}\ \emph {et~al.}(2016)\citenamefont {Shao},
  \citenamefont {Guo},\ and\ \citenamefont {Sandvik}}]{Shao213}%
  \BibitemOpen
  \bibfield  {author} {\bibinfo {author} {\bibfnamefont {H.}~\bibnamefont
  {Shao}}, \bibinfo {author} {\bibfnamefont {W.}~\bibnamefont {Guo}}, \ and\
  \bibinfo {author} {\bibfnamefont {A.~W.}\ \bibnamefont {Sandvik}},\
  }\bibfield  {title} {{\color{Gray}\small \bibinfo {title} {{Quantum
  criticality with two length scales}},\ }}\href {\doibase
  10.1126/science.aad5007} {\bibfield  {journal} {\bibinfo  {journal}
  {Science}\ }\textbf {\bibinfo {volume} {352}},\ \bibinfo {pages} {213}
  (\bibinfo {year} {2016})},\ \Eprint
  {http://arxiv.org/abs/http://science.sciencemag.org/content/352/6282/213.full.pdf}
  {http://science.sciencemag.org/content/352/6282/213.full.pdf} \BibitemShut
  {NoStop}%
\bibitem [{\citenamefont {Hou}\ \emph {et~al.}(2007)\citenamefont {Hou},
  \citenamefont {Chamon},\ and\ \citenamefont {Mudry}}]{PhysRevLett.98.186809}%
  \BibitemOpen
  \bibfield  {author} {\bibinfo {author} {\bibfnamefont {C.-Y.}\ \bibnamefont
  {Hou}}, \bibinfo {author} {\bibfnamefont {C.}~\bibnamefont {Chamon}}, \ and\
  \bibinfo {author} {\bibfnamefont {C.}~\bibnamefont {Mudry}},\ }\bibfield
  {title} {{\color{Gray}\small \bibinfo {title} {{Electron Fractionalization in
  Two-Dimensional Graphenelike Structures}},\ }}\href {\doibase
  10.1103/PhysRevLett.98.186809} {\bibfield  {journal} {\bibinfo  {journal}
  {Phys. Rev. Lett.}\ }\textbf {\bibinfo {volume} {98}},\ \bibinfo {pages}
  {186809} (\bibinfo {year} {2007})}\BibitemShut {NoStop}%
\bibitem [{\citenamefont {Ryu}\ \emph {et~al.}(2009)\citenamefont {Ryu},
  \citenamefont {Mudry}, \citenamefont {Hou},\ and\ \citenamefont
  {Chamon}}]{PhysRevB.80.205319}%
  \BibitemOpen
  \bibfield  {author} {\bibinfo {author} {\bibfnamefont {S.}~\bibnamefont
  {Ryu}}, \bibinfo {author} {\bibfnamefont {C.}~\bibnamefont {Mudry}}, \bibinfo
  {author} {\bibfnamefont {C.-Y.}\ \bibnamefont {Hou}}, \ and\ \bibinfo
  {author} {\bibfnamefont {C.}~\bibnamefont {Chamon}},\ }\bibfield  {title}
  {{\color{Gray}\small \bibinfo {title} {{Masses in graphenelike
  two-dimensional electronic systems: Topological defects in order parameters
  and their fractional exchange statistics}},\ }}\href {\doibase
  10.1103/PhysRevB.80.205319} {\bibfield  {journal} {\bibinfo  {journal} {Phys.
  Rev. B}\ }\textbf {\bibinfo {volume} {80}},\ \bibinfo {pages} {205319}
  (\bibinfo {year} {2009})}\BibitemShut {NoStop}%
\bibitem [{\citenamefont {Roy}\ and\ \citenamefont
  {Herbut}(2010)}]{PhysRevB.82.035429}%
  \BibitemOpen
  \bibfield  {author} {\bibinfo {author} {\bibfnamefont {B.}~\bibnamefont
  {Roy}}\ and\ \bibinfo {author} {\bibfnamefont {I.~F.}\ \bibnamefont
  {Herbut}},\ }\bibfield  {title} {{\color{Gray}\small \bibinfo {title}
  {{Unconventional superconductivity on honeycomb lattice: Theory of Kekule
  order parameter}},\ }}\href {\doibase 10.1103/PhysRevB.82.035429} {\bibfield
  {journal} {\bibinfo  {journal} {Phys. Rev. B}\ }\textbf {\bibinfo {volume}
  {82}},\ \bibinfo {pages} {035429} (\bibinfo {year} {2010})}\BibitemShut
  {NoStop}%
\bibitem [{\citenamefont {Golner}(1973)}]{PhysRevB.8.3419}%
  \BibitemOpen
  \bibfield  {author} {\bibinfo {author} {\bibfnamefont {G.~R.}\ \bibnamefont
  {Golner}},\ }\bibfield  {title} {{\color{Gray}\small \bibinfo {title}
  {{Investigation of the Potts Model Using Renormalization-Group Techniques}},\
  }}\href {\doibase 10.1103/PhysRevB.8.3419} {\bibfield  {journal} {\bibinfo
  {journal} {Phys. Rev. B}\ }\textbf {\bibinfo {volume} {8}},\ \bibinfo {pages}
  {3419} (\bibinfo {year} {1973})}\BibitemShut {NoStop}%
\bibitem [{\citenamefont {Wetterich}(1993)}]{Wetterich:1992yh}%
  \BibitemOpen
  \bibfield  {author} {\bibinfo {author} {\bibfnamefont {C.}~\bibnamefont
  {Wetterich}},\ }\bibfield  {title} {{\color{Gray}\small \bibinfo {title}
  {{Exact evolution equation for the effective potential}},\ }}\href {\doibase
  10.1016/0370-2693(93)90726-X} {\bibfield  {journal} {\bibinfo  {journal}
  {Phys. Lett.}\ }\textbf {\bibinfo {volume} {B301}},\ \bibinfo {pages} {90}
  (\bibinfo {year} {1993})},\ \Eprint {http://arxiv.org/abs/1710.05815}
  {arXiv:1710.05815 [hep-th]} \BibitemShut {NoStop}%
\bibitem [{\citenamefont {Berges}\ \emph {et~al.}(2002)\citenamefont {Berges},
  \citenamefont {Tetradis},\ and\ \citenamefont {Wetterich}}]{Berges:2000ew}%
  \BibitemOpen
  \bibfield  {author} {\bibinfo {author} {\bibfnamefont {J.}~\bibnamefont
  {Berges}}, \bibinfo {author} {\bibfnamefont {N.}~\bibnamefont {Tetradis}}, \
  and\ \bibinfo {author} {\bibfnamefont {C.}~\bibnamefont {Wetterich}},\
  }\bibfield  {title} {{\color{Gray}\small \bibinfo {title} {{Nonperturbative
  renormalization flow in quantum field theory and statistical physics}},\
  }}\href {\doibase 10.1016/S0370-1573(01)00098-9} {\bibfield  {journal}
  {\bibinfo  {journal} {Phys. Rept.}\ }\textbf {\bibinfo {volume} {363}},\
  \bibinfo {pages} {223} (\bibinfo {year} {2002})},\ \Eprint
  {http://arxiv.org/abs/hep-ph/0005122} {arXiv:hep-ph/0005122 [hep-ph]}
  \BibitemShut {NoStop}%
\bibitem [{\citenamefont {Wu}(1982)}]{RevModPhys.54.235}%
  \BibitemOpen
  \bibfield  {author} {\bibinfo {author} {\bibfnamefont {F.~Y.}\ \bibnamefont
  {Wu}},\ }\bibfield  {title} {{\color{Gray}\small \bibinfo {title} {{The Potts
  model}},\ }}\href {\doibase 10.1103/RevModPhys.54.235} {\bibfield  {journal}
  {\bibinfo  {journal} {Rev. Mod. Phys.}\ }\textbf {\bibinfo {volume} {54}},\
  \bibinfo {pages} {235} (\bibinfo {year} {1982})}\BibitemShut {NoStop}%
\bibitem [{\citenamefont {Nelson}(1976)}]{PhysRevB.13.2222}%
  \BibitemOpen
  \bibfield  {author} {\bibinfo {author} {\bibfnamefont {D.~R.}\ \bibnamefont
  {Nelson}},\ }\bibfield  {title} {{\color{Gray}\small \bibinfo {title}
  {{Coexistence-curve singularities in isotropic ferromagnets}},\ }}\href
  {\doibase 10.1103/PhysRevB.13.2222} {\bibfield  {journal} {\bibinfo
  {journal} {Phys. Rev. B}\ }\textbf {\bibinfo {volume} {13}},\ \bibinfo
  {pages} {2222} (\bibinfo {year} {1976})}\BibitemShut {NoStop}%
\bibitem [{\citenamefont {Amit}\ and\ \citenamefont
  {Peliti}(1982)}]{AMIT1982207}%
  \BibitemOpen
  \bibfield  {author} {\bibinfo {author} {\bibfnamefont {D.~J.}\ \bibnamefont
  {Amit}}\ and\ \bibinfo {author} {\bibfnamefont {L.}~\bibnamefont {Peliti}},\
  }\bibfield  {title} {{\color{Gray}\small \bibinfo {title} {On dangerous
  irrelevant operators},\ }}\href {\doibase
  https://doi.org/10.1016/0003-4916(82)90159-2} {\bibfield  {journal} {\bibinfo
   {journal} {Annals of Physics}\ }\textbf {\bibinfo {volume} {140}},\ \bibinfo
  {pages} {207 } (\bibinfo {year} {1982})}\BibitemShut {NoStop}%
\bibitem [{\citenamefont {Novoselov}\ \emph {et~al.}(2005)\citenamefont
  {Novoselov}, \citenamefont {Geim}, \citenamefont {Morozov}, \citenamefont
  {Jiang}, \citenamefont {Katsnelson}, \citenamefont {Grigorieva},
  \citenamefont {Dubonos},\ and\ \citenamefont {Firsov}}]{Novoselov:2005kj}%
  \BibitemOpen
  \bibfield  {author} {\bibinfo {author} {\bibfnamefont {K.~S.}\ \bibnamefont
  {Novoselov}}, \bibinfo {author} {\bibfnamefont {A.~K.}\ \bibnamefont {Geim}},
  \bibinfo {author} {\bibfnamefont {S.~V.}\ \bibnamefont {Morozov}}, \bibinfo
  {author} {\bibfnamefont {D.}~\bibnamefont {Jiang}}, \bibinfo {author}
  {\bibfnamefont {M.~I.}\ \bibnamefont {Katsnelson}}, \bibinfo {author}
  {\bibfnamefont {I.~V.}\ \bibnamefont {Grigorieva}}, \bibinfo {author}
  {\bibfnamefont {S.~V.}\ \bibnamefont {Dubonos}}, \ and\ \bibinfo {author}
  {\bibfnamefont {A.~A.}\ \bibnamefont {Firsov}},\ }\bibfield  {title}
  {{\color{Gray}\small \bibinfo {title} {{Two-dimensional gas of massless Dirac
  fermions in graphene}},\ }}\href {\doibase 10.1038/nature04233} {\bibfield
  {journal} {\bibinfo  {journal} {Nature}\ }\textbf {\bibinfo {volume} {438}},\
  \bibinfo {pages} {197} (\bibinfo {year} {2005})},\ \Eprint
  {http://arxiv.org/abs/cond-mat/0509330} {arXiv:cond-mat/0509330
  [cond-mat.mes-hall]} \BibitemShut {NoStop}%
\bibitem [{\citenamefont {Castro~Neto}\ \emph {et~al.}(2009)\citenamefont
  {Castro~Neto}, \citenamefont {Guinea}, \citenamefont {Peres}, \citenamefont
  {Novoselov},\ and\ \citenamefont {Geim}}]{RevModPhys.81.109}%
  \BibitemOpen
  \bibfield  {author} {\bibinfo {author} {\bibfnamefont {A.~H.}\ \bibnamefont
  {Castro~Neto}}, \bibinfo {author} {\bibfnamefont {F.}~\bibnamefont {Guinea}},
  \bibinfo {author} {\bibfnamefont {N.~M.~R.}\ \bibnamefont {Peres}}, \bibinfo
  {author} {\bibfnamefont {K.~S.}\ \bibnamefont {Novoselov}}, \ and\ \bibinfo
  {author} {\bibfnamefont {A.~K.}\ \bibnamefont {Geim}},\ }\bibfield  {title}
  {{\color{Gray}\small \bibinfo {title} {{The electronic properties of
  graphene}},\ }}\href {\doibase 10.1103/RevModPhys.81.109} {\bibfield
  {journal} {\bibinfo  {journal} {Rev. Mod. Phys.}\ }\textbf {\bibinfo {volume}
  {81}},\ \bibinfo {pages} {109} (\bibinfo {year} {2009})}\BibitemShut
  {NoStop}%
\bibitem [{\citenamefont {Sorella}\ and\ \citenamefont
  {Tosatti}(1992)}]{0295-5075-19-8-007}%
  \BibitemOpen
  \bibfield  {author} {\bibinfo {author} {\bibfnamefont {S.}~\bibnamefont
  {Sorella}}\ and\ \bibinfo {author} {\bibfnamefont {E.}~\bibnamefont
  {Tosatti}},\ }\bibfield  {title} {{\color{Gray}\small \bibinfo {title}
  {{Semi-Metal-Insulator Transition of the Hubbard Model in the Honeycomb
  Lattice}},\ }}\href {http://stacks.iop.org/0295-5075/19/i=8/a=007} {\bibfield
   {journal} {\bibinfo  {journal} {EPL (Europhysics Letters)}\ }\textbf
  {\bibinfo {volume} {19}},\ \bibinfo {pages} {699} (\bibinfo {year}
  {1992})}\BibitemShut {NoStop}%
\bibitem [{\citenamefont {Herbut}(2006)}]{PhysRevLett.97.146401}%
  \BibitemOpen
  \bibfield  {author} {\bibinfo {author} {\bibfnamefont {I.~F.}\ \bibnamefont
  {Herbut}},\ }\bibfield  {title} {{\color{Gray}\small \bibinfo {title}
  {{Interactions and Phase Transitions on Graphene's Honeycomb Lattice}},\
  }}\href {\doibase 10.1103/PhysRevLett.97.146401} {\bibfield  {journal}
  {\bibinfo  {journal} {Phys. Rev. Lett.}\ }\textbf {\bibinfo {volume} {97}},\
  \bibinfo {pages} {146401} (\bibinfo {year} {2006})}\BibitemShut {NoStop}%
\bibitem [{\citenamefont {Honerkamp}(2008)}]{PhysRevLett.100.146404}%
  \BibitemOpen
  \bibfield  {author} {\bibinfo {author} {\bibfnamefont {C.}~\bibnamefont
  {Honerkamp}},\ }\bibfield  {title} {{\color{Gray}\small \bibinfo {title}
  {{Density Waves and Cooper Pairing on the Honeycomb Lattice}},\ }}\href
  {\doibase 10.1103/PhysRevLett.100.146404} {\bibfield  {journal} {\bibinfo
  {journal} {Phys. Rev. Lett.}\ }\textbf {\bibinfo {volume} {100}},\ \bibinfo
  {pages} {146404} (\bibinfo {year} {2008})}\BibitemShut {NoStop}%
\bibitem [{\citenamefont {Raghu}\ \emph {et~al.}(2008)\citenamefont {Raghu},
  \citenamefont {Qi}, \citenamefont {Honerkamp},\ and\ \citenamefont
  {Zhang}}]{PhysRevLett.100.156401}%
  \BibitemOpen
  \bibfield  {author} {\bibinfo {author} {\bibfnamefont {S.}~\bibnamefont
  {Raghu}}, \bibinfo {author} {\bibfnamefont {X.-L.}\ \bibnamefont {Qi}},
  \bibinfo {author} {\bibfnamefont {C.}~\bibnamefont {Honerkamp}}, \ and\
  \bibinfo {author} {\bibfnamefont {S.-C.}\ \bibnamefont {Zhang}},\ }\bibfield
  {title} {{\color{Gray}\small \bibinfo {title} {{Topological Mott
  Insulators}},\ }}\href {\doibase 10.1103/PhysRevLett.100.156401} {\bibfield
  {journal} {\bibinfo  {journal} {Phys. Rev. Lett.}\ }\textbf {\bibinfo
  {volume} {100}},\ \bibinfo {pages} {156401} (\bibinfo {year}
  {2008})}\BibitemShut {NoStop}%
\bibitem [{\citenamefont {Herbut}\ \emph {et~al.}(2009)\citenamefont {Herbut},
  \citenamefont {Juri\ifmmode \check{c}\else \v{c}\fi{}i\ifmmode~\acute{c}\else
  \'{c}\fi{}},\ and\ \citenamefont {Roy}}]{PhysRevB.79.085116}%
  \BibitemOpen
  \bibfield  {author} {\bibinfo {author} {\bibfnamefont {I.~F.}\ \bibnamefont
  {Herbut}}, \bibinfo {author} {\bibfnamefont {V.}~\bibnamefont {Juri\ifmmode
  \check{c}\else \v{c}\fi{}i\ifmmode~\acute{c}\else \'{c}\fi{}}}, \ and\
  \bibinfo {author} {\bibfnamefont {B.}~\bibnamefont {Roy}},\ }\bibfield
  {title} {{\color{Gray}\small \bibinfo {title} {{Theory of interacting
  electrons on the honeycomb lattice}},\ }}\href {\doibase
  10.1103/PhysRevB.79.085116} {\bibfield  {journal} {\bibinfo  {journal} {Phys.
  Rev. B}\ }\textbf {\bibinfo {volume} {79}},\ \bibinfo {pages} {085116}
  (\bibinfo {year} {2009})}\BibitemShut {NoStop}%
\bibitem [{\citenamefont {Weeks}\ and\ \citenamefont
  {Franz}(2010)}]{PhysRevB.81.085105}%
  \BibitemOpen
  \bibfield  {author} {\bibinfo {author} {\bibfnamefont {C.}~\bibnamefont
  {Weeks}}\ and\ \bibinfo {author} {\bibfnamefont {M.}~\bibnamefont {Franz}},\
  }\bibfield  {title} {{\color{Gray}\small \bibinfo {title}
  {{Interaction-driven instabilities of a Dirac semimetal}},\ }}\href {\doibase
  10.1103/PhysRevB.81.085105} {\bibfield  {journal} {\bibinfo  {journal} {Phys.
  Rev. B}\ }\textbf {\bibinfo {volume} {81}},\ \bibinfo {pages} {085105}
  (\bibinfo {year} {2010})}\BibitemShut {NoStop}%
\bibitem [{\citenamefont {de~la Pe\~na}\ \emph {et~al.}(2017)\citenamefont
  {de~la Pe\~na}, \citenamefont {Lichtenstein}, \citenamefont {Honerkamp},\
  and\ \citenamefont {Scherer}}]{PhysRevB.96.205155}%
  \BibitemOpen
  \bibfield  {author} {\bibinfo {author} {\bibfnamefont {D.~S.}\ \bibnamefont
  {de~la Pe\~na}}, \bibinfo {author} {\bibfnamefont {J.}~\bibnamefont
  {Lichtenstein}}, \bibinfo {author} {\bibfnamefont {C.}~\bibnamefont
  {Honerkamp}}, \ and\ \bibinfo {author} {\bibfnamefont {M.~M.}\ \bibnamefont
  {Scherer}},\ }\bibfield  {title} {{\color{Gray}\small \bibinfo {title}
  {{Antiferromagnetism and competing charge instabilities of electrons in
  strained graphene from Coulomb interactions}},\ }}\href {\doibase
  10.1103/PhysRevB.96.205155} {\bibfield  {journal} {\bibinfo  {journal} {Phys.
  Rev. B}\ }\textbf {\bibinfo {volume} {96}},\ \bibinfo {pages} {205155}
  (\bibinfo {year} {2017})}\BibitemShut {NoStop}%
\bibitem [{\citenamefont {Classen}\ \emph {et~al.}(2014)\citenamefont
  {Classen}, \citenamefont {Scherer},\ and\ \citenamefont
  {Honerkamp}}]{PhysRevB.90.035122}%
  \BibitemOpen
  \bibfield  {author} {\bibinfo {author} {\bibfnamefont {L.}~\bibnamefont
  {Classen}}, \bibinfo {author} {\bibfnamefont {M.~M.}\ \bibnamefont
  {Scherer}}, \ and\ \bibinfo {author} {\bibfnamefont {C.}~\bibnamefont
  {Honerkamp}},\ }\bibfield  {title} {{\color{Gray}\small \bibinfo {title}
  {{Instabilities on graphene's honeycomb lattice with electron-phonon
  interactions}},\ }}\href {\doibase 10.1103/PhysRevB.90.035122} {\bibfield
  {journal} {\bibinfo  {journal} {Phys. Rev. B}\ }\textbf {\bibinfo {volume}
  {90}},\ \bibinfo {pages} {035122} (\bibinfo {year} {2014})}\BibitemShut
  {NoStop}%
\bibitem [{\citenamefont {Gomes}\ \emph {et~al.}(2012)\citenamefont {Gomes},
  \citenamefont {Mar}, \citenamefont {Ko}, \citenamefont {Guinea},\ and\
  \citenamefont {Manoharan}}]{Gomes:2012zza}%
  \BibitemOpen
  \bibfield  {author} {\bibinfo {author} {\bibfnamefont {K.~K.}\ \bibnamefont
  {Gomes}}, \bibinfo {author} {\bibfnamefont {W.}~\bibnamefont {Mar}}, \bibinfo
  {author} {\bibfnamefont {W.-H.}\ \bibnamefont {Ko}}, \bibinfo {author}
  {\bibfnamefont {F.}~\bibnamefont {Guinea}}, \ and\ \bibinfo {author}
  {\bibfnamefont {H.~C.}\ \bibnamefont {Manoharan}},\ }\bibfield  {title}
  {{\color{Gray}\small \bibinfo {title} {{Designer Dirac fermions and
  topological phases in molecular graphene}},\ }}\href {\doibase
  10.1038/nature10941} {\bibfield  {journal} {\bibinfo  {journal} {Nature}\
  }\textbf {\bibinfo {volume} {483N7389}},\ \bibinfo {pages} {306} (\bibinfo
  {year} {2012})}\BibitemShut {NoStop}%
\bibitem [{\citenamefont {Guti{\'e}rrez}\ \emph {et~al.}(2016)\citenamefont
  {Guti{\'e}rrez}, \citenamefont {Kim}, \citenamefont {Brown}, \citenamefont
  {Schiros}, \citenamefont {Nordlund}, \citenamefont {Lochocki}, \citenamefont
  {Shen}, \citenamefont {Park},\ and\ \citenamefont
  {Pasupathy}}]{gutierrez2016imaging}%
  \BibitemOpen
  \bibfield  {author} {\bibinfo {author} {\bibfnamefont {C.}~\bibnamefont
  {Guti{\'e}rrez}}, \bibinfo {author} {\bibfnamefont {C.-J.}\ \bibnamefont
  {Kim}}, \bibinfo {author} {\bibfnamefont {L.}~\bibnamefont {Brown}}, \bibinfo
  {author} {\bibfnamefont {T.}~\bibnamefont {Schiros}}, \bibinfo {author}
  {\bibfnamefont {D.}~\bibnamefont {Nordlund}}, \bibinfo {author}
  {\bibfnamefont {E.~B.}\ \bibnamefont {Lochocki}}, \bibinfo {author}
  {\bibfnamefont {K.~M.}\ \bibnamefont {Shen}}, \bibinfo {author}
  {\bibfnamefont {J.}~\bibnamefont {Park}}, \ and\ \bibinfo {author}
  {\bibfnamefont {A.~N.}\ \bibnamefont {Pasupathy}},\ }\bibfield  {title}
  {{\color{Gray}\small \bibinfo {title} {{Imaging chiral symmetry breaking from
  Kekule bond order in graphene}},\ }}\href@noop {} {\bibfield  {journal}
  {\bibinfo  {journal} {Nature Physics}\ }\textbf {\bibinfo {volume} {12}},\
  \bibinfo {pages} {950} (\bibinfo {year} {2016})}\BibitemShut {NoStop}%
\bibitem [{\citenamefont {Semenoff}(1984)}]{PhysRevLett.53.2449}%
  \BibitemOpen
  \bibfield  {author} {\bibinfo {author} {\bibfnamefont {G.~W.}\ \bibnamefont
  {Semenoff}},\ }\bibfield  {title} {{\color{Gray}\small \bibinfo {title}
  {{Condensed-Matter Simulation of a Three-Dimensional Anomaly}},\ }}\href
  {\doibase 10.1103/PhysRevLett.53.2449} {\bibfield  {journal} {\bibinfo
  {journal} {Phys. Rev. Lett.}\ }\textbf {\bibinfo {volume} {53}},\ \bibinfo
  {pages} {2449} (\bibinfo {year} {1984})}\BibitemShut {NoStop}%
\bibitem [{\citenamefont {Gat}\ \emph {et~al.}(1992)\citenamefont {Gat},
  \citenamefont {Kovner},\ and\ \citenamefont {Rosenstein}}]{Gat:1991bf}%
  \BibitemOpen
  \bibfield  {author} {\bibinfo {author} {\bibfnamefont {G.}~\bibnamefont
  {Gat}}, \bibinfo {author} {\bibfnamefont {A.}~\bibnamefont {Kovner}}, \ and\
  \bibinfo {author} {\bibfnamefont {B.}~\bibnamefont {Rosenstein}},\ }\bibfield
   {title} {{\color{Gray}\small \bibinfo {title} {{Chiral phase transitions in
  d = 3 and renormalizability of four Fermi interactions}},\ }}\href {\doibase
  10.1016/0550-3213(92)90095-S} {\bibfield  {journal} {\bibinfo  {journal}
  {Nucl. Phys.}\ }\textbf {\bibinfo {volume} {B385}},\ \bibinfo {pages} {76}
  (\bibinfo {year} {1992})}\BibitemShut {NoStop}%
\bibitem [{\citenamefont {Rosenstein}\ \emph {et~al.}(1993)\citenamefont
  {Rosenstein}, \citenamefont {Yu},\ and\ \citenamefont
  {Kovner}}]{Rosenstein:1993zf}%
  \BibitemOpen
  \bibfield  {author} {\bibinfo {author} {\bibfnamefont {B.}~\bibnamefont
  {Rosenstein}}, \bibinfo {author} {\bibfnamefont {H.-L.}\ \bibnamefont {Yu}},
  \ and\ \bibinfo {author} {\bibfnamefont {A.}~\bibnamefont {Kovner}},\
  }\bibfield  {title} {{\color{Gray}\small \bibinfo {title} {{Critical
  exponents of new universality classes}},\ }}\href {\doibase
  10.1016/0370-2693(93)91253-J} {\bibfield  {journal} {\bibinfo  {journal}
  {Phys. Lett.}\ }\textbf {\bibinfo {volume} {B314}},\ \bibinfo {pages} {381}
  (\bibinfo {year} {1993})}\BibitemShut {NoStop}%
\bibitem [{\citenamefont {Mihaila}\ \emph {et~al.}(2017)\citenamefont
  {Mihaila}, \citenamefont {Zerf}, \citenamefont {Ihrig}, \citenamefont
  {Herbut},\ and\ \citenamefont {Scherer}}]{PhysRevB.96.165133}%
  \BibitemOpen
  \bibfield  {author} {\bibinfo {author} {\bibfnamefont {L.~N.}\ \bibnamefont
  {Mihaila}}, \bibinfo {author} {\bibfnamefont {N.}~\bibnamefont {Zerf}},
  \bibinfo {author} {\bibfnamefont {B.}~\bibnamefont {Ihrig}}, \bibinfo
  {author} {\bibfnamefont {I.~F.}\ \bibnamefont {Herbut}}, \ and\ \bibinfo
  {author} {\bibfnamefont {M.~M.}\ \bibnamefont {Scherer}},\ }\bibfield
  {title} {{\color{Gray}\small \bibinfo {title} {{Gross-Neveu-Yukawa model at
  three loops and Ising critical behavior of Dirac systems}},\ }}\href
  {\doibase 10.1103/PhysRevB.96.165133} {\bibfield  {journal} {\bibinfo
  {journal} {Phys. Rev. B}\ }\textbf {\bibinfo {volume} {96}},\ \bibinfo
  {pages} {165133} (\bibinfo {year} {2017})}\BibitemShut {NoStop}%
\bibitem [{\citenamefont {Iliesiu}\ \emph {et~al.}(2018)\citenamefont
  {Iliesiu}, \citenamefont {Kos}, \citenamefont {Poland}, \citenamefont
  {Pufu},\ and\ \citenamefont {Simmons-Duffin}}]{Iliesiu:2017nrv}%
  \BibitemOpen
  \bibfield  {author} {\bibinfo {author} {\bibfnamefont {L.}~\bibnamefont
  {Iliesiu}}, \bibinfo {author} {\bibfnamefont {F.}~\bibnamefont {Kos}},
  \bibinfo {author} {\bibfnamefont {D.}~\bibnamefont {Poland}}, \bibinfo
  {author} {\bibfnamefont {S.~S.}\ \bibnamefont {Pufu}}, \ and\ \bibinfo
  {author} {\bibfnamefont {D.}~\bibnamefont {Simmons-Duffin}},\ }\bibfield
  {title} {{\color{Gray}\small \bibinfo {title} {{Bootstrapping 3D Fermions
  with Global Symmetries}},\ }}\href {\doibase 10.1007/JHEP01(2018)036}
  {\bibfield  {journal} {\bibinfo  {journal} {JHEP}\ }\textbf {\bibinfo
  {volume} {01}},\ \bibinfo {pages} {036} (\bibinfo {year} {2018})},\ \Eprint
  {http://arxiv.org/abs/1705.03484} {arXiv:1705.03484 [hep-th]} \BibitemShut
  {NoStop}%
\bibitem [{\citenamefont {Zerf}\ \emph {et~al.}(2017)\citenamefont {Zerf},
  \citenamefont {Mihaila}, \citenamefont {Marquard}, \citenamefont {Herbut},\
  and\ \citenamefont {Scherer}}]{PhysRevD.96.096010}%
  \BibitemOpen
  \bibfield  {author} {\bibinfo {author} {\bibfnamefont {N.}~\bibnamefont
  {Zerf}}, \bibinfo {author} {\bibfnamefont {L.~N.}\ \bibnamefont {Mihaila}},
  \bibinfo {author} {\bibfnamefont {P.}~\bibnamefont {Marquard}}, \bibinfo
  {author} {\bibfnamefont {I.~F.}\ \bibnamefont {Herbut}}, \ and\ \bibinfo
  {author} {\bibfnamefont {M.~M.}\ \bibnamefont {Scherer}},\ }\bibfield
  {title} {{\color{Gray}\small \bibinfo {title} {{Four-loop critical exponents
  for the Gross-Neveu-Yukawa models}},\ }}\href {\doibase
  10.1103/PhysRevD.96.096010} {\bibfield  {journal} {\bibinfo  {journal} {Phys.
  Rev. D}\ }\textbf {\bibinfo {volume} {96}},\ \bibinfo {pages} {096010}
  (\bibinfo {year} {2017})}\BibitemShut {NoStop}%
\bibitem [{\citenamefont {Wilson}\ and\ \citenamefont
  {Fisher}(1972)}]{PhysRevLett.28.240}%
  \BibitemOpen
  \bibfield  {author} {\bibinfo {author} {\bibfnamefont {K.~G.}\ \bibnamefont
  {Wilson}}\ and\ \bibinfo {author} {\bibfnamefont {M.~E.}\ \bibnamefont
  {Fisher}},\ }\bibfield  {title} {{\color{Gray}\small \bibinfo {title}
  {{Critical Exponents in 3.99 Dimensions}},\ }}\href {\doibase
  10.1103/PhysRevLett.28.240} {\bibfield  {journal} {\bibinfo  {journal} {Phys.
  Rev. Lett.}\ }\textbf {\bibinfo {volume} {28}},\ \bibinfo {pages} {240}
  (\bibinfo {year} {1972})}\BibitemShut {NoStop}%
\bibitem [{\citenamefont {Zinn-Justin}(2001)}]{ZinnJustin:1999bf}%
  \BibitemOpen
  \bibfield  {author} {\bibinfo {author} {\bibfnamefont {J.}~\bibnamefont
  {Zinn-Justin}},\ }\bibfield  {title} {{\color{Gray}\small \bibinfo {title}
  {{Precise determination of critical exponents and equation of state by field
  theory methods}},\ }}\bibfield  {booktitle} {\emph {\bibinfo {booktitle}
  {{RG-2000: Conference on Renormalization Group Theory at the Turn of the
  Millennium Taxco, Mexico, January 11-15, 1999}}},\ }\href {\doibase
  10.1016/S0370-1573(00)00126-5} {\bibfield  {journal} {\bibinfo  {journal}
  {Phys. Rept.}\ }\textbf {\bibinfo {volume} {344}},\ \bibinfo {pages} {159}
  (\bibinfo {year} {2001})},\ \Eprint {http://arxiv.org/abs/hep-th/0002136}
  {arXiv:hep-th/0002136 [hep-th]} \BibitemShut {NoStop}%
\bibitem [{\citenamefont {Zerf}\ \emph {et~al.}(2016)\citenamefont {Zerf},
  \citenamefont {Lin},\ and\ \citenamefont {Maciejko}}]{Zerf:2016fti}%
  \BibitemOpen
  \bibfield  {author} {\bibinfo {author} {\bibfnamefont {N.}~\bibnamefont
  {Zerf}}, \bibinfo {author} {\bibfnamefont {C.-H.}\ \bibnamefont {Lin}}, \
  and\ \bibinfo {author} {\bibfnamefont {J.}~\bibnamefont {Maciejko}},\
  }\bibfield  {title} {{\color{Gray}\small \bibinfo {title} {{Superconducting
  quantum criticality of topological surface states at three loops}},\ }}\href
  {\doibase 10.1103/PhysRevB.94.205106} {\bibfield  {journal} {\bibinfo
  {journal} {Phys. Rev.}\ }\textbf {\bibinfo {volume} {B94}},\ \bibinfo {pages}
  {205106} (\bibinfo {year} {2016})},\ \Eprint
  {http://arxiv.org/abs/1605.09423} {arXiv:1605.09423 [cond-mat.str-el]}
  \BibitemShut {NoStop}%
\bibitem [{\citenamefont {Rosa}\ \emph {et~al.}(2001)\citenamefont {Rosa},
  \citenamefont {Vitale},\ and\ \citenamefont {Wetterich}}]{Rosa:2000ju}%
  \BibitemOpen
  \bibfield  {author} {\bibinfo {author} {\bibfnamefont {L.}~\bibnamefont
  {Rosa}}, \bibinfo {author} {\bibfnamefont {P.}~\bibnamefont {Vitale}}, \ and\
  \bibinfo {author} {\bibfnamefont {C.}~\bibnamefont {Wetterich}},\ }\bibfield
  {title} {{\color{Gray}\small \bibinfo {title} {{Critical exponents of the
  Gross-Neveu model from the effective average action}},\ }}\href {\doibase
  10.1103/PhysRevLett.86.958} {\bibfield  {journal} {\bibinfo  {journal} {Phys.
  Rev. Lett.}\ }\textbf {\bibinfo {volume} {86}},\ \bibinfo {pages} {958}
  (\bibinfo {year} {2001})},\ \Eprint {http://arxiv.org/abs/hep-th/0007093}
  {arXiv:hep-th/0007093 [hep-th]} \BibitemShut {NoStop}%
\bibitem [{\citenamefont {Hofling}\ \emph {et~al.}(2002)\citenamefont
  {Hofling}, \citenamefont {Nowak},\ and\ \citenamefont
  {Wetterich}}]{Hofling:2002hj}%
  \BibitemOpen
  \bibfield  {author} {\bibinfo {author} {\bibfnamefont {F.}~\bibnamefont
  {Hofling}}, \bibinfo {author} {\bibfnamefont {C.}~\bibnamefont {Nowak}}, \
  and\ \bibinfo {author} {\bibfnamefont {C.}~\bibnamefont {Wetterich}},\
  }\bibfield  {title} {{\color{Gray}\small \bibinfo {title} {{Phase transition
  and critical behavior of the D = 3 Gross-Neveu model}},\ }}\href {\doibase
  10.1103/PhysRevB.66.205111} {\bibfield  {journal} {\bibinfo  {journal} {Phys.
  Rev.}\ }\textbf {\bibinfo {volume} {B66}},\ \bibinfo {pages} {205111}
  (\bibinfo {year} {2002})},\ \Eprint {http://arxiv.org/abs/cond-mat/0203588}
  {arXiv:cond-mat/0203588 [cond-mat]} \BibitemShut {NoStop}%
\bibitem [{\citenamefont {Gies}\ \emph {et~al.}(2010)\citenamefont {Gies},
  \citenamefont {Janssen}, \citenamefont {Rechenberger},\ and\ \citenamefont
  {Scherer}}]{PhysRevD.81.025009}%
  \BibitemOpen
  \bibfield  {author} {\bibinfo {author} {\bibfnamefont {H.}~\bibnamefont
  {Gies}}, \bibinfo {author} {\bibfnamefont {L.}~\bibnamefont {Janssen}},
  \bibinfo {author} {\bibfnamefont {S.}~\bibnamefont {Rechenberger}}, \ and\
  \bibinfo {author} {\bibfnamefont {M.~M.}\ \bibnamefont {Scherer}},\
  }\bibfield  {title} {{\color{Gray}\small \bibinfo {title} {{Phase transition
  and critical behavior of $d=3$ chiral fermion models with left-right
  asymmetry}},\ }}\href {\doibase 10.1103/PhysRevD.81.025009} {\bibfield
  {journal} {\bibinfo  {journal} {Phys. Rev. D}\ }\textbf {\bibinfo {volume}
  {81}},\ \bibinfo {pages} {025009} (\bibinfo {year} {2010})}\BibitemShut
  {NoStop}%
\bibitem [{\citenamefont {Janssen}\ and\ \citenamefont
  {Gies}(2012)}]{Janssen:2012pq}%
  \BibitemOpen
  \bibfield  {author} {\bibinfo {author} {\bibfnamefont {L.}~\bibnamefont
  {Janssen}}\ and\ \bibinfo {author} {\bibfnamefont {H.}~\bibnamefont {Gies}},\
  }\bibfield  {title} {{\color{Gray}\small \bibinfo {title} {{Critical behavior
  of the (2+1)-dimensional Thirring model}},\ }}\href {\doibase
  10.1103/PhysRevD.86.105007} {\bibfield  {journal} {\bibinfo  {journal} {Phys.
  Rev.}\ }\textbf {\bibinfo {volume} {D86}},\ \bibinfo {pages} {105007}
  (\bibinfo {year} {2012})},\ \Eprint {http://arxiv.org/abs/1208.3327}
  {arXiv:1208.3327 [hep-th]} \BibitemShut {NoStop}%
\bibitem [{\citenamefont {Mesterhazy}\ \emph {et~al.}(2012)\citenamefont
  {Mesterhazy}, \citenamefont {Berges},\ and\ \citenamefont {von
  Smekal}}]{Mesterhazy:2012ei}%
  \BibitemOpen
  \bibfield  {author} {\bibinfo {author} {\bibfnamefont {D.}~\bibnamefont
  {Mesterhazy}}, \bibinfo {author} {\bibfnamefont {J.}~\bibnamefont {Berges}},
  \ and\ \bibinfo {author} {\bibfnamefont {L.}~\bibnamefont {von Smekal}},\
  }\bibfield  {title} {{\color{Gray}\small \bibinfo {title} {{Effect of
  short-range interactions on the quantum critical behavior of spinless
  fermions on the honeycomb lattice}},\ }}\href {\doibase
  10.1103/PhysRevB.86.245431} {\bibfield  {journal} {\bibinfo  {journal} {Phys.
  Rev.}\ }\textbf {\bibinfo {volume} {B86}},\ \bibinfo {pages} {245431}
  (\bibinfo {year} {2012})},\ \Eprint {http://arxiv.org/abs/1207.4054}
  {arXiv:1207.4054 [cond-mat.str-el]} \BibitemShut {NoStop}%
\bibitem [{\citenamefont {Janssen}\ and\ \citenamefont
  {Herbut}(2014)}]{Janssen:2014gea}%
  \BibitemOpen
  \bibfield  {author} {\bibinfo {author} {\bibfnamefont {L.}~\bibnamefont
  {Janssen}}\ and\ \bibinfo {author} {\bibfnamefont {I.~F.}\ \bibnamefont
  {Herbut}},\ }\bibfield  {title} {{\color{Gray}\small \bibinfo {title}
  {{Antiferromagnetic critical point on graphene's honeycomb lattice: A
  functional renormalization group approach}},\ }}\href {\doibase
  10.1103/PhysRevB.89.205403} {\bibfield  {journal} {\bibinfo  {journal} {Phys.
  Rev.}\ }\textbf {\bibinfo {volume} {B89}},\ \bibinfo {pages} {205403}
  (\bibinfo {year} {2014})},\ \Eprint {http://arxiv.org/abs/1402.6277}
  {arXiv:1402.6277 [cond-mat.str-el]} \BibitemShut {NoStop}%
\bibitem [{\citenamefont {Classen}\ \emph {et~al.}(2016)\citenamefont
  {Classen}, \citenamefont {Herbut}, \citenamefont {Janssen},\ and\
  \citenamefont {Scherer}}]{PhysRevB.93.125119}%
  \BibitemOpen
  \bibfield  {author} {\bibinfo {author} {\bibfnamefont {L.}~\bibnamefont
  {Classen}}, \bibinfo {author} {\bibfnamefont {I.~F.}\ \bibnamefont {Herbut}},
  \bibinfo {author} {\bibfnamefont {L.}~\bibnamefont {Janssen}}, \ and\
  \bibinfo {author} {\bibfnamefont {M.~M.}\ \bibnamefont {Scherer}},\
  }\bibfield  {title} {{\color{Gray}\small \bibinfo {title} {{Competition of
  density waves and quantum multicritical behavior in Dirac materials from
  functional renormalization}},\ }}\href {\doibase 10.1103/PhysRevB.93.125119}
  {\bibfield  {journal} {\bibinfo  {journal} {Phys. Rev. B}\ }\textbf {\bibinfo
  {volume} {93}},\ \bibinfo {pages} {125119} (\bibinfo {year}
  {2016})}\BibitemShut {NoStop}%
\bibitem [{\citenamefont {Gehring}\ \emph {et~al.}(2015)\citenamefont
  {Gehring}, \citenamefont {Gies},\ and\ \citenamefont
  {Janssen}}]{Gehring:2015vja}%
  \BibitemOpen
  \bibfield  {author} {\bibinfo {author} {\bibfnamefont {F.}~\bibnamefont
  {Gehring}}, \bibinfo {author} {\bibfnamefont {H.}~\bibnamefont {Gies}}, \
  and\ \bibinfo {author} {\bibfnamefont {L.}~\bibnamefont {Janssen}},\
  }\bibfield  {title} {{\color{Gray}\small \bibinfo {title} {{Fixed-point
  structure of low-dimensional relativistic fermion field theories:
  Universality classes and emergent symmetry}},\ }}\href {\doibase
  10.1103/PhysRevD.92.085046} {\bibfield  {journal} {\bibinfo  {journal} {Phys.
  Rev.}\ }\textbf {\bibinfo {volume} {D92}},\ \bibinfo {pages} {085046}
  (\bibinfo {year} {2015})},\ \Eprint {http://arxiv.org/abs/1506.07570}
  {arXiv:1506.07570 [hep-th]} \BibitemShut {NoStop}%
\bibitem [{\citenamefont {Eichhorn}\ \emph {et~al.}(2016)\citenamefont
  {Eichhorn}, \citenamefont {Janssen},\ and\ \citenamefont
  {Scherer}}]{PhysRevD.93.125021}%
  \BibitemOpen
  \bibfield  {author} {\bibinfo {author} {\bibfnamefont {A.}~\bibnamefont
  {Eichhorn}}, \bibinfo {author} {\bibfnamefont {L.}~\bibnamefont {Janssen}}, \
  and\ \bibinfo {author} {\bibfnamefont {M.~M.}\ \bibnamefont {Scherer}},\
  }\bibfield  {title} {{\color{Gray}\small \bibinfo {title} {{Critical $O(N)$
  models above four dimensions: Small-$N$ solutions and stability}},\ }}\href
  {\doibase 10.1103/PhysRevD.93.125021} {\bibfield  {journal} {\bibinfo
  {journal} {Phys. Rev. D}\ }\textbf {\bibinfo {volume} {93}},\ \bibinfo
  {pages} {125021} (\bibinfo {year} {2016})}\BibitemShut {NoStop}%
\bibitem [{\citenamefont {Knorr}(2016)}]{Knorr:2016sfs}%
  \BibitemOpen
  \bibfield  {author} {\bibinfo {author} {\bibfnamefont {B.}~\bibnamefont
  {Knorr}},\ }\bibfield  {title} {{\color{Gray}\small \bibinfo {title} {{Ising
  and Gross-Neveu model in next-to-leading order}},\ }}\href {\doibase
  10.1103/PhysRevB.94.245102} {\bibfield  {journal} {\bibinfo  {journal} {Phys.
  Rev.}\ }\textbf {\bibinfo {volume} {B94}},\ \bibinfo {pages} {245102}
  (\bibinfo {year} {2016})},\ \Eprint {http://arxiv.org/abs/1609.03824}
  {arXiv:1609.03824 [cond-mat.str-el]} \BibitemShut {NoStop}%
\bibitem [{\citenamefont {Knorr}(2017)}]{Knorr:2017yze}%
  \BibitemOpen
  \bibfield  {author} {\bibinfo {author} {\bibfnamefont {B.}~\bibnamefont
  {Knorr}},\ }\bibfield  {title} {{\color{Gray}\small \bibinfo {title}
  {{Critical (Chiral) Heisenberg Model with the Functional Renormalisation
  Group}},\ }}\href@noop {} {\  (\bibinfo {year} {2017})},\ \Eprint
  {http://arxiv.org/abs/1708.06200} {arXiv:1708.06200 [cond-mat.str-el]}
  \BibitemShut {NoStop}%
\bibitem [{\citenamefont {Yin}\ \emph {et~al.}(2017)\citenamefont {Yin},
  \citenamefont {Jian},\ and\ \citenamefont {Yao}}]{Yin:2017gkv}%
  \BibitemOpen
  \bibfield  {author} {\bibinfo {author} {\bibfnamefont {S.}~\bibnamefont
  {Yin}}, \bibinfo {author} {\bibfnamefont {S.-K.}\ \bibnamefont {Jian}}, \
  and\ \bibinfo {author} {\bibfnamefont {H.}~\bibnamefont {Yao}},\ }\bibfield
  {title} {{\color{Gray}\small \bibinfo {title} {{Chiral tricritical point: a
  new universality class in Dirac systems}},\ }}\href@noop {} {\  (\bibinfo
  {year} {2017})},\ \Eprint {http://arxiv.org/abs/1711.10473} {arXiv:1711.10473
  [cond-mat.str-el]} \BibitemShut {NoStop}%
\bibitem [{\citenamefont {Gies}\ \emph {et~al.}(2017)\citenamefont {Gies},
  \citenamefont {Hellwig}, \citenamefont {Wipf},\ and\ \citenamefont
  {Zanusso}}]{Gies:2017tod}%
  \BibitemOpen
  \bibfield  {author} {\bibinfo {author} {\bibfnamefont {H.}~\bibnamefont
  {Gies}}, \bibinfo {author} {\bibfnamefont {T.}~\bibnamefont {Hellwig}},
  \bibinfo {author} {\bibfnamefont {A.}~\bibnamefont {Wipf}}, \ and\ \bibinfo
  {author} {\bibfnamefont {O.}~\bibnamefont {Zanusso}},\ }\bibfield  {title}
  {{\color{Gray}\small \bibinfo {title} {{A functional perspective on emergent
  supersymmetry}},\ }}\href {\doibase 10.1007/JHEP12(2017)132} {\bibfield
  {journal} {\bibinfo  {journal} {JHEP}\ }\textbf {\bibinfo {volume} {12}},\
  \bibinfo {pages} {132} (\bibinfo {year} {2017})},\ \Eprint
  {http://arxiv.org/abs/1705.08312} {arXiv:1705.08312 [hep-th]} \BibitemShut
  {NoStop}%
\bibitem [{\citenamefont {Janssen}\ \emph {et~al.}(2018)\citenamefont
  {Janssen}, \citenamefont {Herbut},\ and\ \citenamefont
  {Scherer}}]{PhysRevB.97.041117}%
  \BibitemOpen
  \bibfield  {author} {\bibinfo {author} {\bibfnamefont {L.}~\bibnamefont
  {Janssen}}, \bibinfo {author} {\bibfnamefont {I.~F.}\ \bibnamefont {Herbut}},
  \ and\ \bibinfo {author} {\bibfnamefont {M.~M.}\ \bibnamefont {Scherer}},\
  }\bibfield  {title} {{\color{Gray}\small \bibinfo {title} {{Compatible orders
  and fermion-induced emergent symmetry in Dirac systems}},\ }}\href {\doibase
  10.1103/PhysRevB.97.041117} {\bibfield  {journal} {\bibinfo  {journal} {Phys.
  Rev. B}\ }\textbf {\bibinfo {volume} {97}},\ \bibinfo {pages} {041117}
  (\bibinfo {year} {2018})}\BibitemShut {NoStop}%
\bibitem [{\citenamefont {Feldmann}\ \emph {et~al.}(2017)\citenamefont
  {Feldmann}, \citenamefont {Wipf},\ and\ \citenamefont
  {Zambelli}}]{Feldmann:2017ooy}%
  \BibitemOpen
  \bibfield  {author} {\bibinfo {author} {\bibfnamefont {P.}~\bibnamefont
  {Feldmann}}, \bibinfo {author} {\bibfnamefont {A.}~\bibnamefont {Wipf}}, \
  and\ \bibinfo {author} {\bibfnamefont {L.}~\bibnamefont {Zambelli}},\
  }\bibfield  {title} {{\color{Gray}\small \bibinfo {title} {{Critical
  Wess-Zumino models with four supercharges from the functional renormalization
  group}},\ }}\href@noop {} {\  (\bibinfo {year} {2017})},\ \Eprint
  {http://arxiv.org/abs/1712.03910} {arXiv:1712.03910 [hep-th]} \BibitemShut
  {NoStop}%
\bibitem [{\citenamefont {Delamotte}\ and\ \citenamefont
  {Canet}(2005)}]{delamotte2005can}%
  \BibitemOpen
  \bibfield  {author} {\bibinfo {author} {\bibfnamefont {B.}~\bibnamefont
  {Delamotte}}\ and\ \bibinfo {author} {\bibfnamefont {L.}~\bibnamefont
  {Canet}},\ }\bibfield  {title} {{\color{Gray}\small \bibinfo {title} {{What
  can be learnt from the nonperturbative renormalization group?}}\ }}\href@noop
  {} {\bibfield  {journal} {\bibinfo  {journal} {Condensed Matter Physics}\ }
  (\bibinfo {year} {2005})}\BibitemShut {NoStop}%
\bibitem [{\citenamefont {Delamotte}\ \emph {et~al.}(2010)\citenamefont
  {Delamotte}, \citenamefont {Dudka}, \citenamefont {Holovatch},\ and\
  \citenamefont {Mouhanna}}]{PhysRevB.82.104432}%
  \BibitemOpen
  \bibfield  {author} {\bibinfo {author} {\bibfnamefont {B.}~\bibnamefont
  {Delamotte}}, \bibinfo {author} {\bibfnamefont {M.}~\bibnamefont {Dudka}},
  \bibinfo {author} {\bibfnamefont {Y.}~\bibnamefont {Holovatch}}, \ and\
  \bibinfo {author} {\bibfnamefont {D.}~\bibnamefont {Mouhanna}},\ }\bibfield
  {title} {{\color{Gray}\small \bibinfo {title} {{Relevance of the fixed
  dimension perturbative approach to frustrated magnets in two and three
  dimensions}},\ }}\href {\doibase 10.1103/PhysRevB.82.104432} {\bibfield
  {journal} {\bibinfo  {journal} {Phys. Rev. B}\ }\textbf {\bibinfo {volume}
  {82}},\ \bibinfo {pages} {104432} (\bibinfo {year} {2010})}\BibitemShut
  {NoStop}%
\bibitem [{\citenamefont {Tissier}\ \emph {et~al.}(2000)\citenamefont
  {Tissier}, \citenamefont {Delamotte},\ and\ \citenamefont
  {Mouhanna}}]{PhysRevLett.84.5208}%
  \BibitemOpen
  \bibfield  {author} {\bibinfo {author} {\bibfnamefont {M.}~\bibnamefont
  {Tissier}}, \bibinfo {author} {\bibfnamefont {B.}~\bibnamefont {Delamotte}},
  \ and\ \bibinfo {author} {\bibfnamefont {D.}~\bibnamefont {Mouhanna}},\
  }\bibfield  {title} {{\color{Gray}\small \bibinfo {title} {{Frustrated
  Heisenberg Magnets: A Nonperturbative Approach}},\ }}\href {\doibase
  10.1103/PhysRevLett.84.5208} {\bibfield  {journal} {\bibinfo  {journal}
  {Phys. Rev. Lett.}\ }\textbf {\bibinfo {volume} {84}},\ \bibinfo {pages}
  {5208} (\bibinfo {year} {2000})}\BibitemShut {NoStop}%
\bibitem [{\citenamefont {Schaefer}\ and\ \citenamefont
  {Wambach}(2005)}]{Schaefer:2004en}%
  \BibitemOpen
  \bibfield  {author} {\bibinfo {author} {\bibfnamefont {B.-J.}\ \bibnamefont
  {Schaefer}}\ and\ \bibinfo {author} {\bibfnamefont {J.}~\bibnamefont
  {Wambach}},\ }\bibfield  {title} {{\color{Gray}\small \bibinfo {title} {{The
  Phase diagram of the quark meson model}},\ }}\href {\doibase
  10.1016/j.nuclphysa.2005.04.012} {\bibfield  {journal} {\bibinfo  {journal}
  {Nucl. Phys.}\ }\textbf {\bibinfo {volume} {A757}},\ \bibinfo {pages} {479}
  (\bibinfo {year} {2005})},\ \Eprint {http://arxiv.org/abs/nucl-th/0403039}
  {arXiv:nucl-th/0403039 [nucl-th]} \BibitemShut {NoStop}%
\bibitem [{\citenamefont {Braun}\ \emph {et~al.}(2010)\citenamefont {Braun},
  \citenamefont {Gies},\ and\ \citenamefont {Pawlowski}}]{Braun:2007bx}%
  \BibitemOpen
  \bibfield  {author} {\bibinfo {author} {\bibfnamefont {J.}~\bibnamefont
  {Braun}}, \bibinfo {author} {\bibfnamefont {H.}~\bibnamefont {Gies}}, \ and\
  \bibinfo {author} {\bibfnamefont {J.~M.}\ \bibnamefont {Pawlowski}},\
  }\bibfield  {title} {{\color{Gray}\small \bibinfo {title} {{Quark Confinement
  from Color Confinement}},\ }}\href {\doibase 10.1016/j.physletb.2010.01.009}
  {\bibfield  {journal} {\bibinfo  {journal} {Phys. Lett.}\ }\textbf {\bibinfo
  {volume} {B684}},\ \bibinfo {pages} {262} (\bibinfo {year} {2010})},\ \Eprint
  {http://arxiv.org/abs/0708.2413} {arXiv:0708.2413 [hep-th]} \BibitemShut
  {NoStop}%
\bibitem [{\citenamefont {Jakubczyk}\ \emph {et~al.}(2009)\citenamefont
  {Jakubczyk}, \citenamefont {Metzner},\ and\ \citenamefont
  {Yamase}}]{PhysRevLett.103.220602}%
  \BibitemOpen
  \bibfield  {author} {\bibinfo {author} {\bibfnamefont {P.}~\bibnamefont
  {Jakubczyk}}, \bibinfo {author} {\bibfnamefont {W.}~\bibnamefont {Metzner}},
  \ and\ \bibinfo {author} {\bibfnamefont {H.}~\bibnamefont {Yamase}},\
  }\bibfield  {title} {{\color{Gray}\small \bibinfo {title} {{Turning a First
  Order Quantum Phase Transition Continuous by Fluctuations: General Flow
  Equations and Application to $d$-Wave Pomeranchuk Instability}},\ }}\href
  {\doibase 10.1103/PhysRevLett.103.220602} {\bibfield  {journal} {\bibinfo
  {journal} {Phys. Rev. Lett.}\ }\textbf {\bibinfo {volume} {103}},\ \bibinfo
  {pages} {220602} (\bibinfo {year} {2009})}\BibitemShut {NoStop}%
\bibitem [{\citenamefont {Pawlowski}\ and\ \citenamefont
  {Rennecke}(2014)}]{Pawlowski:2014zaa}%
  \BibitemOpen
  \bibfield  {author} {\bibinfo {author} {\bibfnamefont {J.~M.}\ \bibnamefont
  {Pawlowski}}\ and\ \bibinfo {author} {\bibfnamefont {F.}~\bibnamefont
  {Rennecke}},\ }\bibfield  {title} {{\color{Gray}\small \bibinfo {title}
  {{Higher order quark-mesonic scattering processes and the phase structure of
  QCD}},\ }}\href {\doibase 10.1103/PhysRevD.90.076002} {\bibfield  {journal}
  {\bibinfo  {journal} {Phys. Rev.}\ }\textbf {\bibinfo {volume} {D90}},\
  \bibinfo {pages} {076002} (\bibinfo {year} {2014})},\ \Eprint
  {http://arxiv.org/abs/1403.1179} {arXiv:1403.1179 [hep-ph]} \BibitemShut
  {NoStop}%
\bibitem [{\citenamefont {Gr\"ater}\ and\ \citenamefont
  {Wetterich}(1995)}]{PhysRevLett.75.378}%
  \BibitemOpen
  \bibfield  {author} {\bibinfo {author} {\bibfnamefont {M.}~\bibnamefont
  {Gr\"ater}}\ and\ \bibinfo {author} {\bibfnamefont {C.}~\bibnamefont
  {Wetterich}},\ }\bibfield  {title} {{\color{Gray}\small \bibinfo {title}
  {{Kosterlitz-Thouless Phase Transition in the Two Dimensional Linear
  $\mathit{\ensuremath{\sigma}}$ Model}},\ }}\href {\doibase
  10.1103/PhysRevLett.75.378} {\bibfield  {journal} {\bibinfo  {journal} {Phys.
  Rev. Lett.}\ }\textbf {\bibinfo {volume} {75}},\ \bibinfo {pages} {378}
  (\bibinfo {year} {1995})}\BibitemShut {NoStop}%
\bibitem [{\citenamefont {Gersdorff}\ and\ \citenamefont
  {Wetterich}(2001)}]{PhysRevB.64.054513}%
  \BibitemOpen
  \bibfield  {author} {\bibinfo {author} {\bibfnamefont {G.~v.}\ \bibnamefont
  {Gersdorff}}\ and\ \bibinfo {author} {\bibfnamefont {C.}~\bibnamefont
  {Wetterich}},\ }\bibfield  {title} {{\color{Gray}\small \bibinfo {title}
  {{Nonperturbative renormalization flow and essential scaling for the
  Kosterlitz-Thouless transition}},\ }}\href {\doibase
  10.1103/PhysRevB.64.054513} {\bibfield  {journal} {\bibinfo  {journal} {Phys.
  Rev. B}\ }\textbf {\bibinfo {volume} {64}},\ \bibinfo {pages} {054513}
  (\bibinfo {year} {2001})}\BibitemShut {NoStop}%
\bibitem [{\citenamefont {Codello}(2012)}]{Codello:2012sc}%
  \BibitemOpen
  \bibfield  {author} {\bibinfo {author} {\bibfnamefont {A.}~\bibnamefont
  {Codello}},\ }\bibfield  {title} {{\color{Gray}\small \bibinfo {title}
  {{Scaling Solutions in Continuous Dimension}},\ }}\href {\doibase
  10.1088/1751-8113/45/46/465006} {\bibfield  {journal} {\bibinfo  {journal}
  {J. Phys.}\ }\textbf {\bibinfo {volume} {A45}},\ \bibinfo {pages} {465006}
  (\bibinfo {year} {2012})},\ \Eprint {http://arxiv.org/abs/1204.3877}
  {arXiv:1204.3877 [hep-th]} \BibitemShut {NoStop}%
\bibitem [{\citenamefont {Codello}\ and\ \citenamefont
  {D'Odorico}(2013)}]{PhysRevLett.110.141601}%
  \BibitemOpen
  \bibfield  {author} {\bibinfo {author} {\bibfnamefont {A.}~\bibnamefont
  {Codello}}\ and\ \bibinfo {author} {\bibfnamefont {G.}~\bibnamefont
  {D'Odorico}},\ }\bibfield  {title} {{\color{Gray}\small \bibinfo {title}
  {{O(N)-Universality Classes and the Mermin-Wagner Theorem}},\ }}\href
  {\doibase 10.1103/PhysRevLett.110.141601} {\bibfield  {journal} {\bibinfo
  {journal} {Phys. Rev. Lett.}\ }\textbf {\bibinfo {volume} {110}},\ \bibinfo
  {pages} {141601} (\bibinfo {year} {2013})}\BibitemShut {NoStop}%
\bibitem [{\citenamefont {Borchardt}\ and\ \citenamefont
  {Eichhorn}(2016)}]{Borchardt:2016kco}%
  \BibitemOpen
  \bibfield  {author} {\bibinfo {author} {\bibfnamefont {J.}~\bibnamefont
  {Borchardt}}\ and\ \bibinfo {author} {\bibfnamefont {A.}~\bibnamefont
  {Eichhorn}},\ }\bibfield  {title} {{\color{Gray}\small \bibinfo {title}
  {{Universal behavior of coupled order parameters below three dimensions}},\
  }}\href {\doibase 10.1103/PhysRevE.94.042105} {\bibfield  {journal} {\bibinfo
   {journal} {Phys. Rev.}\ }\textbf {\bibinfo {volume} {E94}},\ \bibinfo
  {pages} {042105} (\bibinfo {year} {2016})},\ \Eprint
  {http://arxiv.org/abs/1606.07449} {arXiv:1606.07449 [cond-mat.stat-mech]}
  \BibitemShut {NoStop}%
\bibitem [{\citenamefont {Baxter}(1973)}]{Baxter:2000ez}%
  \BibitemOpen
  \bibfield  {author} {\bibinfo {author} {\bibfnamefont {R.~J.}\ \bibnamefont
  {Baxter}},\ }\bibfield  {title} {{\color{Gray}\small \bibinfo {title} {{Potts
  model at critical temperature}},\ }}\href@noop {} {\bibfield  {journal}
  {\bibinfo  {journal} {J. Phys.}\ }\textbf {\bibinfo {volume} {C6}},\ \bibinfo
  {pages} {L445} (\bibinfo {year} {1973})}\BibitemShut {NoStop}%
\bibitem [{\citenamefont {Zinati}\ and\ \citenamefont
  {Codello}(2018)}]{Zinati:2017hdy}%
  \BibitemOpen
  \bibfield  {author} {\bibinfo {author} {\bibfnamefont {R.~B.~A.}\
  \bibnamefont {Zinati}}\ and\ \bibinfo {author} {\bibfnamefont
  {A.}~\bibnamefont {Codello}},\ }\bibfield  {title} {{\color{Gray}\small
  \bibinfo {title} {{Functional RG approach to the Potts model}},\ }}\href
  {\doibase 10.1088/1742-5468/aa9dcc} {\bibfield  {journal} {\bibinfo
  {journal} {J. Stat. Mech.}\ }\textbf {\bibinfo {volume} {1801}},\ \bibinfo
  {pages} {013206} (\bibinfo {year} {2018})},\ \Eprint
  {http://arxiv.org/abs/1707.03410} {arXiv:1707.03410 [cond-mat.stat-mech]}
  \BibitemShut {NoStop}%
\bibitem [{\citenamefont {Ponte}\ and\ \citenamefont
  {Lee}(2014)}]{Ponte:2012ru}%
  \BibitemOpen
  \bibfield  {author} {\bibinfo {author} {\bibfnamefont {P.}~\bibnamefont
  {Ponte}}\ and\ \bibinfo {author} {\bibfnamefont {S.-S.}\ \bibnamefont
  {Lee}},\ }\bibfield  {title} {{\color{Gray}\small \bibinfo {title}
  {{Emergence of supersymmetry on the surface of three dimensional topological
  insulators}},\ }}\href {\doibase 10.1088/1367-2630/16/1/013044} {\bibfield
  {journal} {\bibinfo  {journal} {New J. Phys.}\ }\textbf {\bibinfo {volume}
  {16}},\ \bibinfo {pages} {013044} (\bibinfo {year} {2014})},\ \Eprint
  {http://arxiv.org/abs/1206.2340} {arXiv:1206.2340 [cond-mat.str-el]}
  \BibitemShut {NoStop}%
\bibitem [{\citenamefont {Grover}\ \emph {et~al.}(2014)\citenamefont {Grover},
  \citenamefont {Sheng},\ and\ \citenamefont {Vishwanath}}]{Grover280}%
  \BibitemOpen
  \bibfield  {author} {\bibinfo {author} {\bibfnamefont {T.}~\bibnamefont
  {Grover}}, \bibinfo {author} {\bibfnamefont {D.~N.}\ \bibnamefont {Sheng}}, \
  and\ \bibinfo {author} {\bibfnamefont {A.}~\bibnamefont {Vishwanath}},\
  }\bibfield  {title} {{\color{Gray}\small \bibinfo {title} {{Emergent
  Space-Time Supersymmetry at the Boundary of a Topological Phase}},\ }}\href
  {\doibase 10.1126/science.1248253} {\bibfield  {journal} {\bibinfo  {journal}
  {Science}\ }\textbf {\bibinfo {volume} {344}},\ \bibinfo {pages} {280}
  (\bibinfo {year} {2014})}\BibitemShut {NoStop}%
\bibitem [{\citenamefont {Roy}\ \emph {et~al.}(2016)\citenamefont {Roy},
  \citenamefont {Juricic},\ and\ \citenamefont {Herbut}}]{Roy:2015zna}%
  \BibitemOpen
  \bibfield  {author} {\bibinfo {author} {\bibfnamefont {B.}~\bibnamefont
  {Roy}}, \bibinfo {author} {\bibfnamefont {V.}~\bibnamefont {Juricic}}, \ and\
  \bibinfo {author} {\bibfnamefont {I.~F.}\ \bibnamefont {Herbut}},\ }\bibfield
   {title} {{\color{Gray}\small \bibinfo {title} {{Emergent Lorentz symmetry
  near fermionic quantum critical points in two and three dimensions}},\
  }}\href {\doibase 10.1007/JHEP04(2016)018} {\bibfield  {journal} {\bibinfo
  {journal} {JHEP}\ }\textbf {\bibinfo {volume} {04}},\ \bibinfo {pages} {018}
  (\bibinfo {year} {2016})},\ \Eprint {http://arxiv.org/abs/1510.07650}
  {arXiv:1510.07650 [hep-th]} \BibitemShut {NoStop}%
\bibitem [{\citenamefont {Li}\ \emph {et~al.}(2017)\citenamefont {Li},
  \citenamefont {Vaezi}, \citenamefont {Mendl},\ and\ \citenamefont
  {Yao}}]{Li:2017dkj}%
  \BibitemOpen
  \bibfield  {author} {\bibinfo {author} {\bibfnamefont {Z.-X.}\ \bibnamefont
  {Li}}, \bibinfo {author} {\bibfnamefont {A.}~\bibnamefont {Vaezi}}, \bibinfo
  {author} {\bibfnamefont {C.~B.}\ \bibnamefont {Mendl}}, \ and\ \bibinfo
  {author} {\bibfnamefont {H.}~\bibnamefont {Yao}},\ }\bibfield  {title}
  {{\color{Gray}\small \bibinfo {title} {{Observation of Emergent Spacetime
  Supersymmetry at Superconducting Quantum Criticality}},\ }}\href@noop {} {\
  (\bibinfo {year} {2017})},\ \Eprint {http://arxiv.org/abs/1711.04772}
  {arXiv:1711.04772 [cond-mat.str-el]} \BibitemShut {NoStop}%
\bibitem [{\citenamefont {Litim}(2000)}]{Litim:2000ci}%
  \BibitemOpen
  \bibfield  {author} {\bibinfo {author} {\bibfnamefont {D.~F.}\ \bibnamefont
  {Litim}},\ }\bibfield  {title} {{\color{Gray}\small \bibinfo {title}
  {{Optimization of the exact renormalization group}},\ }}\href {\doibase
  10.1016/S0370-2693(00)00748-6} {\bibfield  {journal} {\bibinfo  {journal}
  {Phys. Lett.}\ }\textbf {\bibinfo {volume} {B486}},\ \bibinfo {pages} {92}
  (\bibinfo {year} {2000})},\ \Eprint {http://arxiv.org/abs/hep-th/0005245}
  {arXiv:hep-th/0005245 [hep-th]} \BibitemShut {NoStop}%
\bibitem [{\citenamefont {Litim}(2001{\natexlab{a}})}]{Litim:2001up}%
  \BibitemOpen
  \bibfield  {author} {\bibinfo {author} {\bibfnamefont {D.~F.}\ \bibnamefont
  {Litim}},\ }\bibfield  {title} {{\color{Gray}\small \bibinfo {title}
  {{Optimized renormalization group flows}},\ }}\href {\doibase
  10.1103/PhysRevD.64.105007} {\bibfield  {journal} {\bibinfo  {journal} {Phys.
  Rev.}\ }\textbf {\bibinfo {volume} {D64}},\ \bibinfo {pages} {105007}
  (\bibinfo {year} {2001}{\natexlab{a}})},\ \Eprint
  {http://arxiv.org/abs/hep-th/0103195} {arXiv:hep-th/0103195 [hep-th]}
  \BibitemShut {NoStop}%
\bibitem [{\citenamefont {Litim}(2001{\natexlab{b}})}]{Litim:2001fd}%
  \BibitemOpen
  \bibfield  {author} {\bibinfo {author} {\bibfnamefont {D.~F.}\ \bibnamefont
  {Litim}},\ }\bibfield  {title} {{\color{Gray}\small \bibinfo {title} {{Mind
  the gap}},\ }}\bibfield  {booktitle} {\emph {\bibinfo {booktitle} {{The exact
  renormalization group. Proceedings, 2nd Conference, Rome, Italy, September
  18-22, 2000}}},\ }\href {\doibase 10.1142/S0217751X01004748} {\bibfield
  {journal} {\bibinfo  {journal} {Int. J. Mod. Phys.}\ }\textbf {\bibinfo
  {volume} {A16}},\ \bibinfo {pages} {2081} (\bibinfo {year}
  {2001}{\natexlab{b}})},\ \Eprint {http://arxiv.org/abs/hep-th/0104221}
  {arXiv:hep-th/0104221 [hep-th]} \BibitemShut {NoStop}%
\bibitem [{\citenamefont {Litim}(2002)}]{Litim:2002cf}%
  \BibitemOpen
  \bibfield  {author} {\bibinfo {author} {\bibfnamefont {D.~F.}\ \bibnamefont
  {Litim}},\ }\bibfield  {title} {{\color{Gray}\small \bibinfo {title}
  {{Critical exponents from optimized renormalization group flows}},\ }}\href
  {\doibase 10.1016/S0550-3213(02)00186-4} {\bibfield  {journal} {\bibinfo
  {journal} {Nucl. Phys.}\ }\textbf {\bibinfo {volume} {B631}},\ \bibinfo
  {pages} {128} (\bibinfo {year} {2002})},\ \Eprint
  {http://arxiv.org/abs/hep-th/0203006} {arXiv:hep-th/0203006 [hep-th]}
  \BibitemShut {NoStop}%
\bibitem [{\citenamefont {Pawlowski}\ \emph {et~al.}(2017)\citenamefont
  {Pawlowski}, \citenamefont {Scherer}, \citenamefont {Schmidt},\ and\
  \citenamefont {Wetzel}}]{PAWLOWSKI2017165}%
  \BibitemOpen
  \bibfield  {author} {\bibinfo {author} {\bibfnamefont {J.~M.}\ \bibnamefont
  {Pawlowski}}, \bibinfo {author} {\bibfnamefont {M.~M.}\ \bibnamefont
  {Scherer}}, \bibinfo {author} {\bibfnamefont {R.}~\bibnamefont {Schmidt}}, \
  and\ \bibinfo {author} {\bibfnamefont {S.~J.}\ \bibnamefont {Wetzel}},\
  }\bibfield  {title} {{\color{Gray}\small \bibinfo {title} {Physics and the
  choice of regulators in functional renormalisation group flows},\ }}\href
  {\doibase https://doi.org/10.1016/j.aop.2017.06.017} {\bibfield  {journal}
  {\bibinfo  {journal} {Annals of Physics}\ }\textbf {\bibinfo {volume}
  {384}},\ \bibinfo {pages} {165 } (\bibinfo {year} {2017})}\BibitemShut
  {NoStop}%
\bibitem [{\citenamefont {Borchardt}\ and\ \citenamefont
  {Knorr}(2015)}]{Borchardt:2015rxa}%
  \BibitemOpen
  \bibfield  {author} {\bibinfo {author} {\bibfnamefont {J.}~\bibnamefont
  {Borchardt}}\ and\ \bibinfo {author} {\bibfnamefont {B.}~\bibnamefont
  {Knorr}},\ }\bibfield  {title} {{\color{Gray}\small \bibinfo {title} {{Global
  solutions of functional fixed point equations via pseudospectral methods}},\
  }}\href {\doibase 10.1103/PhysRevD.93.089904, 10.1103/PhysRevD.91.105011}
  {\bibfield  {journal} {\bibinfo  {journal} {Phys. Rev.}\ }\textbf {\bibinfo
  {volume} {D91}},\ \bibinfo {pages} {105011} (\bibinfo {year} {2015})},\
  \bibinfo {note} {[Erratum: Phys. Rev.D93,no.8,089904(2016)]},\ \Eprint
  {http://arxiv.org/abs/1502.07511} {arXiv:1502.07511 [hep-th]} \BibitemShut
  {NoStop}%
\bibitem [{\citenamefont {Borchardt}\ and\ \citenamefont
  {Knorr}(2016)}]{Borchardt:2016pif}%
  \BibitemOpen
  \bibfield  {author} {\bibinfo {author} {\bibfnamefont {J.}~\bibnamefont
  {Borchardt}}\ and\ \bibinfo {author} {\bibfnamefont {B.}~\bibnamefont
  {Knorr}},\ }\bibfield  {title} {{\color{Gray}\small \bibinfo {title}
  {{Solving functional flow equations with pseudo-spectral methods}},\ }}\href
  {\doibase 10.1103/PhysRevD.94.025027} {\bibfield  {journal} {\bibinfo
  {journal} {Phys. Rev.}\ }\textbf {\bibinfo {volume} {D94}},\ \bibinfo {pages}
  {025027} (\bibinfo {year} {2016})},\ \Eprint
  {http://arxiv.org/abs/1603.06726} {arXiv:1603.06726 [hep-th]} \BibitemShut
  {NoStop}%
\end{thebibliography}%
\end{document}